\pdfoutput=1

\documentclass[12pt,a4paper]{article}

\usepackage{ifthen} 
\usepackage{booktabs,dcolumn}
\usepackage{tabularx}
\usepackage{verbatim}
\newboolean{pdflatex}
\setboolean{pdflatex}{true} 

\newboolean{articletitles}
\setboolean{articletitles}{true} 

\newboolean{uprightparticles}
\setboolean{uprightparticles}{false} 

\newboolean{inbibliography}
\setboolean{inbibliography}{false} 

\def\paperauthors{LHCb collaboration} 
\def\paperasciititle{Measurement of branching fractions of charmless four-body Lb and Xib decays} 
\def\papertitle{Measurement of branching fractions of charmless four-body \Lb and \Xibz decays} 
\def\paperkeywords{{High Energy Physics}, {LHCb}} 
\def\papercopyright{CERN on behalf of the LHCb collaboration}
\def\paperlicence{CC-BY-4.0}
\def\paperlicenceurl{https://creativecommons.org/licenses/by/4.0/}

\usepackage[top=1in, bottom=1.25in, left=1in, right=1in]{geometry}

\usepackage{microtype}
\usepackage{lineno}  
\usepackage{xspace} 
\usepackage{caption} 

\usepackage{graphicx}  
\usepackage{color}
\usepackage{colortbl}
\graphicspath{{./figs/}} 

\usepackage{amsmath} 
\usepackage{amssymb}
\usepackage{amsfonts}
\usepackage{upgreek} 

\newcommand*\patchAmsMathEnvironmentForLineno[1]{%
\expandafter\let\csname old#1\expandafter\endcsname\csname #1\endcsname
\expandafter\let\csname oldend#1\expandafter\endcsname\csname
end#1\endcsname
 \renewenvironment{#1}%
   {\linenomath\csname old#1\endcsname}%
   {\csname oldend#1\endcsname\endlinenomath}%
}
\newcommand*\patchBothAmsMathEnvironmentsForLineno[1]{%
  \patchAmsMathEnvironmentForLineno{#1}%
  \patchAmsMathEnvironmentForLineno{#1*}%
}
\AtBeginDocument{%
\patchBothAmsMathEnvironmentsForLineno{equation}%
\patchBothAmsMathEnvironmentsForLineno{align}%
\patchBothAmsMathEnvironmentsForLineno{flalign}%
\patchBothAmsMathEnvironmentsForLineno{alignat}%
\patchBothAmsMathEnvironmentsForLineno{gather}%
\patchBothAmsMathEnvironmentsForLineno{multline}%
\patchBothAmsMathEnvironmentsForLineno{eqnarray}%
}

\usepackage{hyperxmp}

\usepackage[pdftex,
            pdfauthor={\paperauthors},
            pdftitle={\paperasciititle},
            pdfkeywords={\paperkeywords},
            pdfcopyright={Copyright (C) \papercopyright},
            pdflicenseurl={\paperlicenceurl}]{hyperref}

\usepackage[all]{hypcap} 

\usepackage{xspace} 
\usepackage{upgreek}

\def\lhcb {\mbox{LHCb}\xspace}

\def\MagUp {\mbox{\em Mag\kern -0.05em Up}\xspace}

\ifthenelse{\boolean{uprightparticles}}%
{
 
 \def\Pgamma      {\ensuremath{\upgamma}\xspace}

 \def\Peta        {\ensuremath{\upeta}\xspace}

 \def\Ppi         {\ensuremath{\uppi}\xspace}

 \def\Pchi        {\ensuremath{\upchi}\xspace}                 
 \def\Ppsi        {\ensuremath{\uppsi}\xspace}

 \def\PDelta      {\ensuremath{\Delta}\xspace}                 
 \def\PXi      {\ensuremath{\Xi}\xspace}                 
 \def\PLambda      {\ensuremath{\Lambda}\xspace}                 
 \def\PSigma      {\ensuremath{\Sigma}\xspace}                 
 \def\POmega      {\ensuremath{\Omega}\xspace}                 
 \def\PUpsilon      {\ensuremath{\Upsilon}\xspace}                 
 
 \def\PB      {\ensuremath{\mathrm{B}}\xspace}                 
                  
 \def\PD      {\ensuremath{\mathrm{D}}\xspace}

 \def\PJ      {\ensuremath{\mathrm{J}}\xspace}                 
 \def\PK      {\ensuremath{\mathrm{K}}\xspace}

 \def\PX      {\ensuremath{\mathrm{X}}\xspace}

 \def\Pb      {\ensuremath{\mathrm{b}}\xspace}                 
 \def\Pc      {\ensuremath{\mathrm{c}}\xspace}                 
 \def\Pd      {\ensuremath{\mathrm{d}}\xspace}

 \def\Ph      {\ensuremath{\mathrm{h}}\xspace}                 
 \def\Pi      {\ensuremath{\mathrm{i}}\xspace}

 \def\Pp      {\ensuremath{\mathrm{p}}\xspace}

 \def\Ps      {\ensuremath{\mathrm{s}}\xspace}                 
                  
 \def\Pu      {\ensuremath{\mathrm{u}}\xspace}

}
{
 
 \def\Pgamma      {\ensuremath{\gamma}\xspace}

 \def\Peta        {\ensuremath{\eta}\xspace}

 \def\Ppi         {\ensuremath{\pi}\xspace}

 \def\Pchi        {\ensuremath{\chi}\xspace}                 
 \def\Ppsi        {\ensuremath{\psi}\xspace}                 
                  
 \mathchardef\PDelta="7101
 \mathchardef\PXi="7104
 \mathchardef\PLambda="7103
 \mathchardef\PSigma="7106
 \mathchardef\POmega="710A
 \mathchardef\PUpsilon="7107
                  
 \def\PB      {\ensuremath{B}\xspace}                 
                  
 \def\PD      {\ensuremath{D}\xspace}

 \def\PJ      {\ensuremath{J}\xspace}                 
 \def\PK      {\ensuremath{K}\xspace}

 \def\PX      {\ensuremath{X}\xspace}

 \def\Pb      {\ensuremath{b}\xspace}                 
 \def\Pc      {\ensuremath{c}\xspace}                 
 \def\Pd      {\ensuremath{d}\xspace}

 \def\Ph      {\ensuremath{h}\xspace}                 
 \def\Pi      {\ensuremath{i}\xspace}

 \def\Pp      {\ensuremath{p}\xspace}

 \def\Ps      {\ensuremath{s}\xspace}                 
                  
 \def\Pu      {\ensuremath{u}\xspace}

}

\makeatletter
\ifcase \@ptsize \relax
  \newcommand{\miniscule}{\@setfontsize\miniscule{4}{5}}
\or
  \newcommand{\miniscule}{\@setfontsize\miniscule{5}{6}}
\or
  \newcommand{\miniscule}{\@setfontsize\miniscule{5}{6}}
\fi
\makeatother

\DeclareRobustCommand{\optbar}[1]{\shortstack{{\miniscule (\rule[.5ex]{1.25em}{.18mm})}
  \\ [-.7ex] $#1$}}

\def\g      {{\ensuremath{\Pgamma}}\xspace}

\def\uquark    {{\ensuremath{\Pu}}\xspace}

\def\dquark    {{\ensuremath{\Pd}}\xspace}

\def\squark    {{\ensuremath{\Ps}}\xspace}

\def\cquark    {{\ensuremath{\Pc}}\xspace}
\def\cquarkbar {{\ensuremath{\overline \cquark}}\xspace}

\def\bquark    {{\ensuremath{\Pb}}\xspace}

\def\pion   {{\ensuremath{\Ppi}}\xspace}

\def\pip    {{\ensuremath{\pion^+}}\xspace}
\def\pim    {{\ensuremath{\pion^-}}\xspace}

\def\pimp   {{\ensuremath{\pion^\mp}}\xspace}

\def\kaon    {{\ensuremath{\PK}}\xspace}
  \def\Kbar    {{\kern 0.2em\overline{\kern -0.2em \PK}{}}\xspace}

\def\KorKbar    {\kern 0.18em\optbar{\kern -0.18em K}{}\xspace}

\def\Kp      {{\ensuremath{\kaon^+}}\xspace}
\def\Km      {{\ensuremath{\kaon^-}}\xspace}
\def\Kpm     {{\ensuremath{\kaon^\pm}}\xspace}

\def\KS      {{\ensuremath{\kaon^0_{\mathrm{ \scriptscriptstyle S}}}}\xspace}

\newcommand{\etapr}{\ensuremath{\Peta^{\prime}}\xspace}

  \def\Dbar    {{\kern 0.2em\overline{\kern -0.2em \PD}{}}\xspace}
\def\D       {{\ensuremath{\PD}}\xspace}

\def\DorDbar    {\kern 0.18em\optbar{\kern -0.18em D}{}\xspace}
\def\Dz      {{\ensuremath{\D^0}}\xspace}

\def\Dp      {{\ensuremath{\D^+}}\xspace}

\def\Dsp     {{\ensuremath{\D^+_\squark}}\xspace}

\def\B       {{\ensuremath{\PB}}\xspace}
\def\Bbar    {{\ensuremath{\kern 0.18em\overline{\kern -0.18em \PB}{}}}\xspace}

\def\BorBbar    {\kern 0.18em\optbar{\kern -0.18em B}{}\xspace}

\def\Bd      {{\ensuremath{\B^0}}\xspace}
\def\Bs      {{\ensuremath{\B^0_\squark}}\xspace}

\def\jpsi     {{\ensuremath{{\PJ\mskip -3mu/\mskip -2mu\Ppsi\mskip 2mu}}}\xspace}

\def\chiczero {{\ensuremath{\Pchi_{\cquark 0}}}\xspace}

  \def\Y#1S{\ensuremath{\PUpsilon{(#1S)}}\xspace}

\def\Xires       {{\ensuremath{\PXi}}\xspace}

\def\Lz          {{\ensuremath{\PLambda}}\xspace}
\def\Lbar        {{\ensuremath{\kern 0.1em\overline{\kern -0.1em\PLambda}}}\xspace}
\def\LorLbar    {\kern 0.18em\optbar{\kern -0.18em \PLambda}{}\xspace}

\def\Lb      {{\ensuremath{\Lz^0_\bquark}}\xspace}

\def\Lc      {{\ensuremath{\Lz^+_\cquark}}\xspace}

\def\Xibz    {{\ensuremath{\Xires^0_\bquark}}\xspace}

\def\Xicp    {{\ensuremath{\Xires^+_\cquark}}\xspace}

\def\BF         {{\ensuremath{\mathcal{B}}}\xspace}

\newcommand{\decay}[2]{\ensuremath{#1\!\to #2}\xspace}         

\def\to                 {\ensuremath{\rightarrow}\xspace}

\def\eps   {{\ensuremath{\varepsilon}}\xspace}

\def\CP                {{\ensuremath{C\!P}}\xspace}

\def\AT#1     {\ensuremath{A_{\mathrm{T}}^{#1}}\xspace}           

\def\C#1      {\ensuremath{\mathcal{C}_{#1}}\xspace}                       
\def\Cp#1     {\ensuremath{\mathcal{C}_{#1}^{'}}\xspace}                    
\def\Ceff#1   {\ensuremath{\mathcal{C}_{#1}^{\mathrm{(eff)}}}\xspace}        
\def\Cpeff#1  {\ensuremath{\mathcal{C}_{#1}^{'\mathrm{(eff)}}}\xspace}       
\def\Ope#1    {\ensuremath{\mathcal{O}_{#1}}\xspace}                       
\def\Opep#1   {\ensuremath{\mathcal{O}_{#1}^{'}}\xspace}                    

\newcommand{\tev}{\ifthenelse{\boolean{inbibliography}}{\ensuremath{~T\kern -0.05em eV}}{\ensuremath{\mathrm{\,Te\kern -0.1em V}}}\xspace}
\newcommand{\gev}{\ensuremath{\mathrm{\,Ge\kern -0.1em V}}\xspace}
\newcommand{\mev}{\ensuremath{\mathrm{\,Me\kern -0.1em V}}\xspace}
\newcommand{\kev}{\ensuremath{\mathrm{\,ke\kern -0.1em V}}\xspace}
\newcommand{\ev}{\ensuremath{\mathrm{\,e\kern -0.1em V}}\xspace}
\newcommand{\gevc}{\ensuremath{{\mathrm{\,Ge\kern -0.1em V\!/}c}}\xspace}
\newcommand{\mevc}{\ensuremath{{\mathrm{\,Me\kern -0.1em V\!/}c}}\xspace}
\newcommand{\gevcc}{\ensuremath{{\mathrm{\,Ge\kern -0.1em V\!/}c^2}}\xspace}
\newcommand{\gevgevcccc}{\ensuremath{{\mathrm{\,Ge\kern -0.1em V^2\!/}c^4}}\xspace}
\newcommand{\mevcc}{\ensuremath{{\mathrm{\,Me\kern -0.1em V\!/}c^2}}\xspace}

\def\mum  {\ensuremath{{\,\upmu\mathrm{m}}}\xspace}

\def\invfb   {\ensuremath{\mbox{\,fb}^{-1}}\xspace}

\newcommand{\chisq}{\ensuremath{\chi^2}\xspace}

\newcommand{\chisqip}{\ensuremath{\chi^2_{\text{IP}}}\xspace}

\newcommand{\chisqvtx}{\ensuremath{\chi^2_{\text{vtx}}}\xspace}

\def\gsim{{~\raise.15em\hbox{$>$}\kern-.85em
          \lower.35em\hbox{$\sim$}~}\xspace}
\def\lsim{{~\raise.15em\hbox{$<$}\kern-.85em
          \lower.35em\hbox{$\sim$}~}\xspace}

\def\ptot       {\mbox{$p$}\xspace}
\def\pt         {\mbox{$p_{\mathrm{ T}}$}\xspace}

\def\rad{\ensuremath{\mathrm{ \,rad}}\xspace}

\def\evtgen     {\mbox{\textsc{EvtGen}}\xspace}

\def\geant      {\mbox{\textsc{Geant4}}\xspace}

\def\photos     {\mbox{\textsc{Photos}}\xspace}

\def\pythia     {\mbox{\textsc{Pythia}}\xspace}

\def\tell1  {TELL1\xspace}
\def\ukl1   {UKL1\xspace}

\newcommand{\eg}{\mbox{\itshape e.g.}\xspace}
\newcommand{\ie}{\mbox{\itshape i.e.}\xspace}

\usepackage{cite} 
\usepackage{mciteplus}

\def\LbToppipipi              {\ensuremath{\Lb\to \Pp\Ppi^-\Ppi^+\Ppi^-}\xspace}
\def\LbTopKpipi               {\ensuremath{\Lb\to \Pp\PK^-\Ppi^+\Ppi^-}\xspace}

\def\LbTopKKpi                {\ensuremath{\Lb\to \Pp\PK^-\PK^+\Ppi^-}\xspace}

\def\LbTopKKK                 {\ensuremath{\Lb\to \Pp\PK^-\PK^+\PK^-}\xspace}
\def\XibzTopKpipi             {\ensuremath{\Xibz\to \Pp\PK^-\Ppi^+\Ppi^-}\xspace}

\def\XibzTopKpiK              {\ensuremath{\Xibz\to \Pp\PK^-\Ppi^+\PK^-}\xspace}
\def\XibzTopKKK               {\ensuremath{\Xibz\to \Pp\PK^-\PK^+\PK^-}\xspace}

\def\XbzTophhhz               {\ensuremath{X_b \to \Pp\Ph\Ph'\Ph''}\xspace}

\def\Xb                       {\ensuremath{\PX^0_\bquark}\xspace}

\def\pipi                     {\ensuremath{\Ppi\Ppi}\xspace}

\def\kk                       {\ensuremath{\PK\PK}\xspace}

\def\ppipipi                  {\ensuremath{\Pp\Ppi^-\Ppi^+\Ppi^-}\xspace}
\def\pKpipi                   {\ensuremath{\Pp\PK^-\Ppi^+\Ppi^-}\xspace}

\def\pKKpi                    {\ensuremath{\Pp\PK^-\PK^+\Ppi^-}\xspace}
\def\pKpiK                    {\ensuremath{\Pp\PK^-\Ppi^+\PK^-}\xspace}

\def\pKKK                     {\ensuremath{\Pp\PK^-\PK^+\PK^-}\xspace}

\def\phhh                     {\ensuremath{\Pp\Ph\Ph\Ph}\xspace}
\def\phhhpiz                  {\ensuremath{\Pp\Ph\Ph\Ph\Ppi^0}\xspace}

\def\chisqfd                  {\ensuremath{\chi^2_{\rm FD}}\xspace}

\def\LbToLcpiLcTopKpi         {\ensuremath{\Lb\to(\Lc\to\Pp\PK^-\Ppi^+)\Ppi^-}\xspace}
\def\LbToLcpi                 {\ensuremath{\Lb\to\Lc\Ppi^-}\xspace}

\def\LbTopKetap               {\ensuremath{\Lb\to\Pp\PK^-\etapr}\xspace}
\def\LbToppietap              {\ensuremath{\Lb\to\Pp\Ppi^-\etapr}\xspace}
\def\EtapTopipig              {\ensuremath{\etapr\to\Ppi^+\Ppi^-\gamma}\xspace}
\def\LbTopKpipipiz            {\ensuremath{\Lb\to\Pp\PK^-\Ppi^+\Ppi^-\Ppi^0}\xspace}
\def\LbToppipipipiz           {\ensuremath{\Lb\to\Pp\Ppi^-\Ppi^+\Ppi^-\Ppi^0}\xspace}
\def\LbTopKKpipiz             {\ensuremath{\Lb\to\Pp\PK^-\PK^+\Ppi^-\Ppi^0}\xspace}
\def\LbTopKKKpiz              {\ensuremath{\Lb\to\Pp\PK^-\PK^+\PK^-\Ppi^0}\xspace}
\def\LbTopKKKg                {\ensuremath{\Lb\to\Pp\PK^-\PK^+\PK^-\g}\xspace}

\def\DzorDzbar                {\kern 0.18em\optbar{\kern -0.18em \Dz}{}\xspace}

\def\LcTopKpi                 {\ensuremath{\Lc\to\Pp\PK^-\Ppi^+}\xspace}

\def\pid                      {\ensuremath{\mathrm{PID}}\xspace}

\def\Y                        {\ensuremath{\mathcal{N}}\xspace}

\ifthenelse{\boolean{uprightparticles}}%
{
 
}
{
 
}

\def\had  {\ensuremath{\Ph}\xspace}

\def\BdorBdbar    {\ensuremath{\kern 0.18em\optbar{\kern -0.18em B}{}^0}\xspace}
\def\BsorBsbar    {\ensuremath{\kern 0.18em\optbar{\kern  0.06em B_s}{}^0}\xspace}

\def\pipi  {\ensuremath{\pion^+\pion^-}\xspace}

\def\kk      {\ensuremath{\Kp\Km}\xspace}

\def\BdtoKsKK   {\decay{\Bd}{\KS \Kp \Km}}
\def\BdtoKsPiPi   {\decay{\Bd}{\KS \pip \pim}}

\def\BdtoKsKpPim   {\decay{\Bd}{\KS \Kp \pim}}
\def\BdtoKsPipKm   {\decay{\Bd}{\KS \Km \pip}}

\def\BstoKsKK   {\decay{\Bs}{\KS \Kp \Km}}
\def\BstoKsPiPi   {\decay{\Bs}{\KS \pip \pim}}

\def\BstoKsKpPim   {\decay{\Bs}{\KS \Kp \pim}}
\def\BstoKsPipKm   {\decay{\Bs}{\KS \Km \pip}}

\def\KsPiPi{\ensuremath{\KS \pip \pim}\xspace}

\def\KsKK{\ensuremath{\KS \Kp \Km}\xspace}

\def\KsKpPim{\ensuremath{\KS \Kp \pim}\xspace}
\def\KsPipKm{\ensuremath{\KS \pip \Km}\xspace}

\def\LL   {Long-Long\xspace}
\def\DD   {Down-Down\xspace}

\def \fitCombShape{Polynomial}
\def \fitCombModel{from5150}

\newcommand{\plotIfExists}[2]{
  \IfFileExists{#1}{\includegraphics[width=#2\textwidth]{#1}}{\includegraphics[width=#2\textwidth]{#1}}
}    
\newcommand{\plotOne}[2]{
  \ifthenelse{\boolean{pdflatex}}{
    \plotIfExists{#1.pdf}{#2}
  }{
    \plotIfExists{#1.eps}{#2}
  }
}

\newcommand{\plotDataFitResults}[3]{ 
  \begin{figure}[p]
    \begin{center}
      \ifthenelse{\equal{\fitCombShape}{Exponential}}
                 {\def \tempFitCombShape{Exponential}}
                 {\def \tempFitCombShape{Polynomial}}
                 
                 \plotOne{figs/FitResults/#1/\fitCombModel-Louis-\tempFitCombShape-StrongPcut-#1-KSKK#2_#3-Standard-DoubleCB}{0.35}
                 \plotOne{figs/FitResults/#1/\fitCombModel-Louis-\tempFitCombShape-StrongPcut-#1-KSKK#2_#3-Standard-DoubleCB_log}{0.35}\\
                 \plotOne{figs/FitResults/#1/\fitCombModel-Louis-\tempFitCombShape-StrongPcut-#1-KSKpi#2_#3-Standard-DoubleCB}{0.35}
                 \plotOne{figs/FitResults/#1/\fitCombModel-Louis-\tempFitCombShape-StrongPcut-#1-KSKpi#2_#3-Standard-DoubleCB_log}{0.35}\\
                 \plotOne{figs/FitResults/#1/\fitCombModel-Louis-\tempFitCombShape-StrongPcut-#1-KSpiK#2_#3-Standard-DoubleCB}{0.35}
                 \plotOne{figs/FitResults/#1/\fitCombModel-Louis-\tempFitCombShape-StrongPcut-#1-KSpiK#2_#3-Standard-DoubleCB_log}{0.35}\\
                 \plotOne{figs/FitResults/#1/\fitCombModel-Louis-\tempFitCombShape-StrongPcut-#1-KSpipi#2_#3-Standard-DoubleCB}{0.35}
                 \plotOne{figs/FitResults/#1/\fitCombModel-Louis-\tempFitCombShape-StrongPcut-#1-KSpipi#2_#3-Standard-DoubleCB_log}{0.35}\\
                 \ifthenelse{\equal{#2}{DD}}{
                   \caption{Results of the simultaneous fit to data (\DD, #3) with the \MakeLowercase{#1} BDT optimisation. The modes \KsKK, \KsKpPim, \KsPipKm and \KsPiPi are shown from top to bottom. The left-hand side plots show the results with a linear scale and the right-hand side with a logarithmic scale.}
                 }{
                   \caption{Results of the simultaneous fit to data (\LL, #3) with the \MakeLowercase{#1} BDT optimisation. The modes \KsKK, \KsKpPim, \KsPipKm and \KsPiPi are shown from top to bottom. The left-hand side plots show the results with a linear scale and the right-hand side with a logarithmic scale.}
                 }
                 \label{fig:FitResult:#1:#2:#3}
    \end{center}
  \end{figure}
}
\newcommand{\plotSignalMCFitResults}[3]{
  \begin{figure}[!htbp]
    \begin{center}
      \plotOne{figs/FitModel/Signal/#1/Bd2KSKK#2_#3-MCFit-Louis-DoubleCB-Standard_log}{0.35}
      \plotOne{figs/FitModel/Signal/#1/Bs2KSKK#2_#3-MCFit-Louis-DoubleCB-Standard_log}{0.35}\\
      \plotOne{figs/FitModel/Signal/#1/Bd2KSKpi#2_#3-MCFit-Louis-DoubleCB-Standard_log}{0.35}
      \plotOne{figs/FitModel/Signal/#1/Bs2KSKpi#2_#3-MCFit-Louis-DoubleCB-Standard_log}{0.35}\\
      \plotOne{figs/FitModel/Signal/#1/Bd2KSpiK#2_#3-MCFit-Louis-DoubleCB-Standard_log}{0.35}
      \plotOne{figs/FitModel/Signal/#1/Bs2KSpiK#2_#3-MCFit-Louis-DoubleCB-Standard_log}{0.35}\\
      \plotOne{figs/FitModel/Signal/#1/Bd2KSpipi#2_#3-MCFit-Louis-DoubleCB-Standard_log}{0.35}
      \plotOne{figs/FitModel/Signal/#1/Bs2KSpipi#2_#3-MCFit-Louis-DoubleCB-Standard_log}{0.35}\\
      \ifthenelse{\equal{#2}{DD}}{
        \caption{Result of the simultaneous fit to simulated samples of the signal decays for \DD \KS reconstruction mode, using the \MakeLowercase{#1} optimisation of the BDT, and shown using a logarithmic scale. \KsKK, \KsKpPim, \KsPipKm, and \KsPiPi are shown from top to bottom, while \Bd decays are shown on the left and \Bs decays on the right. }
      }
                 {
                   \caption{Result of the simultaneous fit to simulated samples of the signal decays for \LL \KS reconstruction mode, using the \MakeLowercase{#1} optimisation of the BDT, and shown using a logarithmic scale. \KsKK, \KsKpPim, \KsPipKm, and \KsPiPi are shown from top to bottom, while \Bd decays are shown on the left and \Bs decays on the right. }
                 }
                 \label{fig:FitModel:Signal:#1:#2:#3}
    \end{center}
\end{figure}
}

\newcommand{\plotSplots}[3]{
  \ifthenelse{\equal{#3}{log}}
             {\def\suffix{FitStandard_log}}
             {\def\suffix{FitStandard}}
             \ifthenelse{\equal{#3}{log}}
                        {\def\scale{logarithmic}\xspace}
                        {\def\scale{linear}\xspace}
                        \begin{figure}[!htbp]
                          \begin{center}
                            \plotOne{figs/FitResults/sWeights/#1/sWeights-from5150-Louis-PolSlopes-StrongPcut-#1-KSKKDD_#2_\suffix}{0.35}
                            \plotOne{figs/FitResults/sWeights/#1/sWeights-from5150-Louis-PolSlopes-StrongPcut-#1-KSKKLL_#2_\suffix}{0.35}\\
                            \plotOne{figs/FitResults/sWeights/#1/sWeights-from5150-Louis-PolSlopes-StrongPcut-#1-KSKpiDD_#2_\suffix}{0.35}
                            \plotOne{figs/FitResults/sWeights/#1/sWeights-from5150-Louis-PolSlopes-StrongPcut-#1-KSKpiLL_#2_\suffix}{0.35}\\
                            \plotOne{figs/FitResults/sWeights/#1/sWeights-from5150-Louis-PolSlopes-StrongPcut-#1-KSpiKDD_#2_\suffix}{0.35}
                            \plotOne{figs/FitResults/sWeights/#1/sWeights-from5150-Louis-PolSlopes-StrongPcut-#1-KSpiKLL_#2_\suffix}{0.35}\\
                            \plotOne{figs/FitResults/sWeights/#1/sWeights-from5150-Louis-PolSlopes-StrongPcut-#1-KSpipiDD_#2_\suffix}{0.35}
                            \plotOne{figs/FitResults/sWeights/#1/sWeights-from5150-Louis-PolSlopes-StrongPcut-#1-KSpipiLL_#2_\suffix}{0.35}\\
                            \caption{Result of the invariant mass fits used for the sWeights extraction with the \MakeLowercase{#1} BDT optimisation on #2 data (\scale~scale). \KsKK, \KsKpPim, \KsPipKm, and \KsPiPi are shown from top to bottom, \DD on the left, \LL on the right.}
                            \label{fig:FitResult:Splots:#1:#2:#3}
                          \end{center}
                        \end{figure}
}
\newcommand{\plotSplotsDalitz}[3]{ 
  \ifthenelse{\equal{\fitCombShape}{Exponential}}
             {\def \tempFitCombShape{Exponential}}
             {\def \tempFitCombShape{PolSlopes}}
             
             \begin{figure}[!htbp]
               \begin{center}
                 \plotOne{figs/FitResults/sWeights/#1/sWeights-\fitCombModel-Louis-\tempFitCombShape-StrongPcut-#1-KSKKDD_#3_#2_Dalitz}{0.35}
                 \plotOne{figs/FitResults/sWeights/#1/sWeights-\fitCombModel-Louis-\tempFitCombShape-StrongPcut-#1-KSKKLL_#3_#2_Dalitz}{0.35}\\
                 \plotOne{figs/FitResults/sWeights/#1/sWeights-\fitCombModel-Louis-\tempFitCombShape-StrongPcut-#1-KSKpiDD_#3_#2_Dalitz}{0.35}
                 \plotOne{figs/FitResults/sWeights/#1/sWeights-\fitCombModel-Louis-\tempFitCombShape-StrongPcut-#1-KSKpiLL_#3_#2_Dalitz}{0.35}\\
                 \plotOne{figs/FitResults/sWeights/#1/sWeights-\fitCombModel-Louis-\tempFitCombShape-StrongPcut-#1-KSpiKDD_#3_#2_Dalitz}{0.35}
                 \plotOne{figs/FitResults/sWeights/#1/sWeights-\fitCombModel-Louis-\tempFitCombShape-StrongPcut-#1-KSpiKLL_#3_#2_Dalitz}{0.35}\\
                 \plotOne{figs/FitResults/sWeights/#1/sWeights-\fitCombModel-Louis-\tempFitCombShape-StrongPcut-#1-KSpipiDD_#3_#2_Dalitz}{0.35}
                 \plotOne{figs/FitResults/sWeights/#1/sWeights-\fitCombModel-Louis-\tempFitCombShape-StrongPcut-#1-KSpipiLL_#3_#2_Dalitz}{0.35}\\
                 \ifthenelse{\equal{#2}{Bd}}{
                   \caption{Distribution of \Bd signal sWeights extracted with the \MakeLowercase{#1} BDT optimisation in #3 data. \KsKpPim, \KsPipKm, and \KsPiPi are shown from top to bottom, \DD on the left, \LL on the right. The corrections due to the presence of species with fixed yields is not applied here.}
                 }{
                   \caption{Distribution of \Bs signal sWeights extracted with the \MakeLowercase{#1} BDT optimisation in #3 data. \KsKpPim, \KsPipKm, and \KsPiPi are shown from top to bottom, \DD on the left, \LL on the right. The corrections due to the presence of species with fixed yields is not applied here.}
                 } 
                 \label{fig:FitResult:Splots:#1:#3:Dalitz#2}
               \end{center}
             \end{figure}
}

\newcommand{\makeSystTabular}[2]{
  \begin{table}[!htbp]
    \begin{center}
      \ifthenelse{\equal{#2}{KK}}
                 {       
                   \caption{Systematic uncertainties originating from the fit model of the \KsKK modes (using \MakeLowercase{#1} BDT optimisation). The numbers are rounded to the upper integer value, except for the total yield. The total systematic error is defined as the sum in quadrature of all the components.}
                 }{}
                 \ifthenelse{\equal{#2}{Kpi}}
                            {
                              \caption{Systematic uncertainties originating from the fit model of the \KsKpPim modes (using \MakeLowercase{#1} BDT optimisation). The numbers are rounded to the upper integer value, except for the total yield. The total systematic error is defined as the sum in quadrature of all the components.}
                            }{}
                            \ifthenelse{\equal{#2}{piK}}        
                                       {
                                         \caption{Systematic uncertainties originating from the fit model of the \KsPipKm modes (using \MakeLowercase{#1} BDT optimisation). The numbers are rounded to the upper integer value, except for the total yield. The total systematic error is defined as the sum in quadrature of all the components.}
                                       }{}
                                       \ifthenelse{\equal{#2}{pipi}}
                                                  {
                                                    \caption{Systematic uncertainties originating from the fit model of the \KsPiPi modes (using \MakeLowercase{#1} BDT optimisation). The numbers are rounded to the upper integer value, except for the total yield. The total systematic error is defined as the sum in quadrature of all the components.}
                                                  }{}
                                                  \label{Table:FitSyst:#1:#2}
                                                  \resizebox{\textwidth}{!}{
                                                    \input{tables/SystI-PolSlopes-from5150-#1-#2.txt}
                                                  }
    \end{center}
  \end{table}
  
}

\newcommand{\makeInvMass}[1]{
  \ifthenelse{\equal{#1}{Bd2KSpipi} \or \equal{#1}{Bs2KSpipi}}              {\def\myInvMass {pipi}}
             {\ifthenelse{\equal{#1}{Bd2KSpiK} \or \equal{#1}{Bs2KSpiK}}    {\def\myInvMass {piK}}
               {\ifthenelse{\equal{#1}{Bd2KSKpi} \or \equal{#1}{Bs2KSKpi}}  {\def\myInvMass {Kpi}}
                 {\ifthenelse{\equal{#1}{Bd2KSKK} \or \equal{#1}{Bs2KSKK}}  {\def\myInvMass {KK}}{\def\myInvMass ERROR}}
               }
             }

}
\newcommand{\makeMode}[1]{
  \ifthenelse{\equal{#1}{Bd2KSpipi}}                     {\def\myMode {\BdtoKsPiPi}}
             {\ifthenelse{\equal{#1}{Bd2KSpiK}}          {\def\myMode {\BdtoKsPipKm}}
               {\ifthenelse{\equal{#1}{Bd2KSKpi}}        {\def\myMode {\BdtoKsKpPim}}
                 {\ifthenelse{\equal{#1}{Bd2KSKK}}       {\def\myMode {\BdtoKsKK}}
                   {\ifthenelse{\equal{#1}{Bs2KSpipi}}           {\def\myMode {\BstoKsPiPi}}
                     {\ifthenelse{\equal{#1}{Bs2KSpiK}}          {\def\myMode {\BstoKsPipKm}}
                       {\ifthenelse{\equal{#1}{Bs2KSKpi}}        {\def\myMode {\BstoKsKpPim}}
                         {\ifthenelse{\equal{#1}{Bs2KSKK}}       {\def\myMode {\BstoKsKK}}{\def\myMode{ERROR}}
                         }
                       }
                     }
                   }
                 }
               }
             }
}

\newcommand{\plotGeomEff}[3]{
  \makeInvMass{#2}
  \makeMode{#2}
  \ifthenelse{\equal{#1}{2011}}{\def\redYear {2011}}{\def\redYear {2012}}
  \centering
  \includegraphics[width=0.45\textwidth]{figs/EfficiencyMaps/#1/Geometry/NoSel/#2/Efficiency_Map_NoSel_#2_#3_#1_\myInvMass.pdf.eps}
  \includegraphics[width=0.45\textwidth]{figs/EfficiencyMaps/#1/Geometry/NoSel/#2/Spline_Eff_NoSel_#2_#3_#1_\myInvMass.pdf.eps}
  \includegraphics[width=0.45\textwidth]{figs/EfficiencyMaps/#1/Geometry/NoSel/#2/Efficiency_errorHi_Map_NoSel_#2_#3_#1_\myInvMass.pdf.eps}
  \includegraphics[width=0.45\textwidth]{figs/EfficiencyMaps/#1/Geometry/NoSel/#2/Efficiency_errorLo_Map_NoSel_#2_#3_#1_\myInvMass.pdf.eps}
  \caption{
    (Top) $\epsilon^{\rm geom}$ as a function of the \myMode
    square Dalitz plot position obtained from \redYear-conditions generator-level signal MC:
    (left) the raw histogram,
    (right) smoothed using a 2D cubic spline.
    (Bottom) the (left) upper and (right) lower uncertainties on the histogram bins.
    Uncertainties are due to MC statistics.
  }
  \label{fig:geoeff:GeoEff:#1:#2:#3}
}

\newcommand{\plotTrackCorr}[4]{
  \makeInvMass{#2}
  \makeMode{#2}
  \centering
  \includegraphics[width=0.45\textwidth]{figs/EfficiencyMaps/#1/Tracking/#4/#2/Trk_AllEff-#4-#2-#3-#1-\myInvMass.pdf.eps}
  \includegraphics[width=0.45\textwidth]{figs/EfficiencyMaps/#1/Tracking/#4/#2/Spline_Eff_#4_#2_#3_#1_\myInvMass.pdf.eps}\\
  \includegraphics[width=0.45\textwidth]{figs/EfficiencyMaps/#1/Tracking/#4/#2/Tracking_errorHi_bootstrap_#4_#2_#3_\myInvMass_#1.pdf.eps}
  \includegraphics[width=0.45\textwidth]{figs/EfficiencyMaps/#1/Tracking/#4/#2/Tracking_errorHi_bootstrap_#4_#2_#3_\myInvMass_#1.pdf.eps}
  \caption{
    (Top) Combined tracking efficiency corrections in the \myMode signal mode for #3 and #1 configuration:
    (left) the raw histogram obtained from MC simulation and (right) smoothed using a 2D cubic spline.  
    (Bottom) the (left) upper and (right) lower uncertainties on the histogram bins.
  }
  \label{fig:DPefficiency-trk-all-#1-#2-#3-#4}
}

\newcommand{\plotTrackCorrMode}[1]{
  \begin{figure}[!htb]
    \plotTrackCorr{2011}{#1}{DD}{Loose}
  \end{figure}
  \begin{figure}[!htb]
    \plotTrackCorr{2011}{#1}{LL}{Loose}
  \end{figure}
}

\newcommand{\plotTrigCorr}[4]{
  \makeInvMass{#1}
  \makeMode{#1}
  \ifthenelse{\equal{#4}{TOS}}{
    \def\trigComment {$\epsilon^{\rm L0TOS|sel\&geom}_{\rm data}/\epsilon^{\rm L0TOS|sel\&geom}_{\rm MC}$}}{
    \def\trigComment {$\epsilon^{\rm !L0TOS|sel\&geom}_{\rm data}/\epsilon^{\rm !L0TOS|sel\&geom}_{\rm MC}$}}
  
  \centering
  \includegraphics[width=0.45\textwidth]{figs/EfficiencyMaps/2011/L0#4/#3/#1/L0#4-Correction-#3-#1-#2-2011-\myInvMass.pdf.eps}
  \includegraphics[width=0.45\textwidth]{figs/EfficiencyMaps/2011/L0#4/#3/#1/Spline_Eff_#3_#1_#2_2011_\myInvMass.pdf.eps}\\
  \includegraphics[width=0.45\textwidth]{figs/EfficiencyMaps/2011/L0#4/#3/#1/L0#4_errorHi_combined_#3_#1_#2_\myInvMass_2011.pdf.eps}
  \includegraphics[width=0.45\textwidth]{figs/EfficiencyMaps/2011/L0#4/#3/#1/L0#4_errorLo_combined_#3_#1_#2_\myInvMass_2011.pdf.eps}
  \caption{
    (Top) \trigComment across the \myMode #2 square Dalitz plot (2011+2012 combined):
    (left) the raw histogram obtained using the procedure described in the text and (right) smoothed using a 2D cubic spline.  
    (Bottom) the (left) upper and (right) lower uncertainties on the histogram bins.
  }
  \label{fig:DPcorrection-#1-#2-#3-#4}
}

\newcommand{\plotSelEff}[5]{
  \makeInvMass{#2}
  \makeMode{#2}
  \ifthenelse{\equal{#5}{TOS}}{\def\selComment {{\tt L0Hadron\_TOS}}}{\def\selComment {{\tt L0Global\_TIS\&\&!L0Hadron\_TOS}}}
  \centering
  \includegraphics[width=0.45\textwidth]{figs/EfficiencyMaps/#1/Sel#5/#4/#2/#2-Eff-#4-#2-#3-#1-\myInvMass.pdf.eps}
  \includegraphics[width=0.45\textwidth]{figs/EfficiencyMaps/#1/Sel#5/#4/#2/Spline_Eff_#4_#2_#3_#1_\myInvMass.pdf.eps}\\
  \includegraphics[width=0.45\textwidth]{figs/EfficiencyMaps/#1/Sel#5/#4/#2/#2-ErrorHi-#4-#2-#3-#1-\myInvMass.pdf.eps}
  \includegraphics[width=0.45\textwidth]{figs/EfficiencyMaps/#1/Sel#5/#4/#2/#2-ErrorLo-#4-#2-#3-#1-\myInvMass.pdf.eps}
  \caption{
    (Top) $\epsilon^{\rm sel|geom}$ across the \myMode #3 #1 square Dalitz plot for \selComment \MakeLowercase{#4} BDT candidates:
    (left) the raw histogram obtained using the procedure described in the text and (right) smoothed using a 2D cubic spline.  
    (Bottom) the (left) upper and (right) lower uncertainties on the histogram bins.
  }
  \label{fig:DPefficiency-sel-#1-#2-#3-#4-#5}
}

\newcommand{\plotSelEffMode}[2]{
  \begin{figure}[!htb]
    \plotSelEff{2011}{#1}{DD}{#1}{TOS}
  \end{figure}
  \begin{figure}[!htb]
    \plotSelEff{2011}{#1}{DD}{#1}{TIS}
  \end{figure}
  \begin{figure}[!htb]
    \plotSelEff{2011}{#1}{LL}{#1}{TOS}
  \end{figure}
  \begin{figure}[!htb]
    \plotSelEff{2011}{#1}{LL}{#1}{TIS}
  \end{figure}
}

\usepackage{longtable} 
\usepackage{multirow}

\begin{document}

\renewcommand{\thefootnote}{\fnsymbol{footnote}}
\setcounter{footnote}{1}


\begin{titlepage}
\pagenumbering{roman}

\vspace*{-1.5cm}
\centerline{\large EUROPEAN ORGANIZATION FOR NUCLEAR RESEARCH (CERN)}
\vspace*{1.5cm}
\noindent
\begin{tabular*}{\linewidth}{lc@{\extracolsep{\fill}}r@{\extracolsep{0pt}}}
\ifthenelse{\boolean{pdflatex}}
{\vspace*{-1.5cm}\mbox{\!\!\!\includegraphics[width=.14\textwidth]{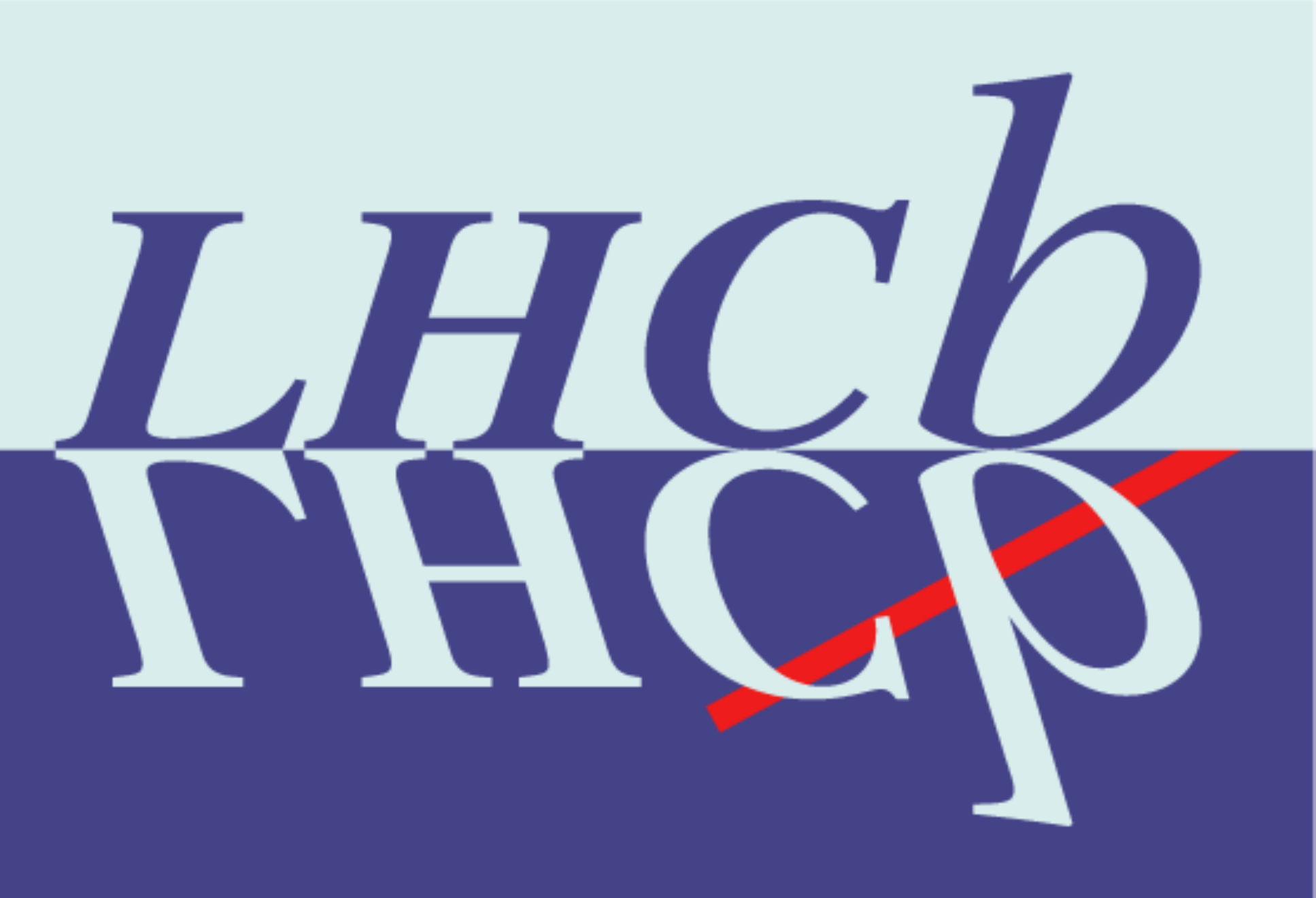}} & &}%
{\vspace*{-1.2cm}\mbox{\!\!\!\includegraphics[width=.12\textwidth]{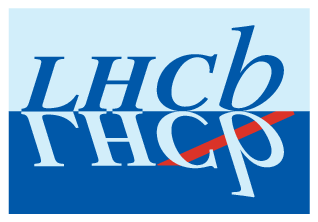}} & &}%
\\ 
 & & CERN-EP-2017-278 \\  
 & & LHCb-PAPER-2017-034 \\  
 & & February 28, 2018 \\ 
 & & \\

\end{tabular*}

\vspace*{4.0cm}

{\normalfont\bfseries\boldmath\huge
\begin{center}
  \papertitle 
\end{center}
}

\vspace*{2.0cm}

\begin{center}
\paperauthors\footnote{Authors are listed at the end of this paper.}
\end{center}

\vspace{\fill}

\begin{abstract}
  \noindent
  A search for charmless four-body decays of \Lb and \Xibz baryons with a proton and three charged mesons (either kaons or pions)
 in the final state is performed. The data sample used was recorded in 2011 and 2012 with the LHCb experiment and corresponds to an integrated luminosity of 3 \invfb. Six decay modes are observed, among which \LbTopKpipi, \LbTopKKK, \XibzTopKpipi and \XibzTopKpiK are established for the first time. Their branching fractions (including the ratio of hadronisation fractions in the case of the \Xibz baryon) are determined relative to the \LbToLcpi decay.
  
\end{abstract}

\vspace*{2.0cm}

\begin{center}
  Published in JHEP 02 (2018) 098
\end{center}

\vspace{\fill}

{\footnotesize 
\centerline{\copyright~\papercopyright, licence \href{\paperlicenceurl}{\paperlicence}.}}
\vspace*{2mm}

\end{titlepage}


\newpage
\setcounter{page}{2}
\mbox{~}
%
%
%
%

\cleardoublepage


\renewcommand{\thefootnote}{\arabic{footnote}}
\setcounter{footnote}{0}



\pagestyle{plain} 
\setcounter{page}{1}
\pagenumbering{arabic}


%

\section{Introduction}
\label{sec:introduction}

The abundant production of  \Lb and \Xibz baryons in proton-proton collisions at the Large Hadron Collider (LHC) gives the \lhcb experiment the opportunity to study multibody charmless weak decays of \bquark-flavoured baryons. 
The establishment of \Lb and \Xibz baryon signals will allow the measurements of their branching fractions as well as the \CP-violating asymmetries in their decay.

The measurements of \CP-violation phenomena present, so far, a consistent interpretation within the Standard Model paradigm~\cite{CKMfitter2015}.  Nonvanishing \CP-violating asymmetries have been observed in the decays of both $K$ and \B mesons~\cite{PDG2017}. In contrast, \CP violation has not been clearly observed in baryon decays although evidence for nonvanishing \CP asymmetries in \bquark-flavoured baryon decays has been recently reported by the \lhcb collaboration~\cite{LHCb-PAPER-2016-030}.

{\it A priori} relevant decay modes to observe \CP violation in \bquark-baryon decays are multibody charmless decays that can proceed simultaneously through the charged-current $\bquark \to \uquark$ transition or the neutral-current $\bquark \to \squark,\dquark$ transitions. The resulting interference exhibits a weak-phase difference. Furthermore, the charmless multibody decays of \bquark baryons contain  rich resonance structures, both in the low-mass two-body baryon resonances (\ie the $\Pp\PK^-$, $\Pp\Ppi^-$ and $\Pp\Ppi^+$ invariant mass spectra) and in the two-body nonbaryonic resonances (\ie the \pipi, \Kpm\pimp and \kk invariant mass spectra). Consequently, \CP asymmetries might receive significant enhancement from the strong-phase differences coming from the interference of these resonances. Taken together, these factors make multibody charmless \bquark-baryon decays well suited for a potential first observation of \CP violation in the baryon sector. Conversely, the presence of nonpredictible strong phases makes a potential observation of \CP violation difficult to interpret in terms of the weak phase of the Cabibbo-Kobayashi-Maskawa (CKM) quark-mixing matrix~\cite{PhysRevLett.10.531, Kobayashi:1973fv}. 

This work focuses on a study of seven decays,\footnote{Charge conjugation is implied throughout this document, unless stated otherwise.} namely \LbToppipipi, \LbTopKpipi, \LbTopKKpi, 
\LbTopKKK, \XibzTopKpipi, \XibzTopKpiK and \XibzTopKKK, defining five exclusive final states to study.  The signal candidates are fully reconstructed and selected by means of optimised particle identification and topological criteria. A simultaneous fit to the invariant mass distribution of the candidates in the five experimental spectra is performed to determine the signal yields.
The branching fractions, relative to the well-known normalisation channel \LbToLcpiLcTopKpi~\cite{LHCb-PAPER-2014-004}, are subsequently determined.

\section{Detector and data set}
\label{sec:Detector}

The analysis reported here is performed using $pp$ collision data
recorded with the LHCb detector, corresponding to an integrated luminosity of
1.0\invfb at a centre-of-mass energy of 7\tev in 2011 and 2.0\invfb at a
centre-of-mass energy of 8\tev in 2012.
The \lhcb detector~\cite{Alves:2008zz,LHCb-DP-2014-002} is a single-arm forward
spectrometer covering the \mbox{pseudorapidity} range $2<\eta <5$,
designed for the study of particles containing \bquark or \cquark
quarks. The detector includes a high-precision tracking system
consisting of a silicon-strip vertex detector surrounding the $pp$
interaction region,
a large-area silicon-strip detector located
upstream of a dipole magnet with a bending power of about
$4{\mathrm{\,Tm}}$, and three stations of silicon-strip detectors and straw
drift tubes
placed downstream of the magnet.
The tracking system provides a measurement of momentum, \ptot, of charged particles with
a relative uncertainty that varies from 0.5\% at low momentum to 1.0\% at 200\gevc.
The minimum distance of a track to a primary vertex (PV), the impact parameter (IP), 
is measured with a resolution of $(15+29/\pt)\mum$,
where \pt is the component of the momentum transverse to the beam, in\,\gevc.
Different types of charged hadrons are distinguished using information
from two ring-imaging Cherenkov detectors.
Photons, electrons and hadrons are identified by a calorimeter system consisting of
scintillating-pad and preshower detectors, an electromagnetic
calorimeter and a hadronic calorimeter. Muons are identified by a
system composed of alternating layers of iron and multiwire
proportional chambers.

Simulated data samples are used to investigate backgrounds from other
\bquark-hadron decays and also to study the detection and reconstruction
efficiencies of the signals.
In the simulation, $pp$ collisions are generated using
\pythia~\cite{Sjostrand:2006za,*Sjostrand:2007gs} with a specific \lhcb
configuration~\cite{LHCb-PROC-2010-056}.
Decays of hadronic particles are described by \evtgen~\cite{Lange:2001uf}
in which final-state radiation is generated using
\photos~\cite{Golonka:2005pn}.
The interactions of the generated particles with the detector, and its
response, are implemented using the \geant toolkit~\cite{Allison:2006ve,
*Agostinelli:2002hh} as described in Ref.~\cite{LHCb-PROC-2011-006}.

\section{Trigger and event selection}
\label{sec:Selection}

The online event selection is performed by a trigger\cite{LHCb-DP-2012-004} 
that consists of a hardware stage, based on information from the calorimeter and muon systems, followed 
by a software stage, in which all charged particles with $\pt>500\,(300)\mevc$ are reconstructed for 2011\,(2012) data.
At the hardware trigger stage, events are required to have a muon with high \pt or a hadron, photon or electron with 
high transverse energy. The software 
trigger requires a two-, three- or four-track secondary vertex with a significant displacement from all primary $pp$ 
interaction vertices. At least one charged particle must have a transverse momentum $\pt > 1.7\,(1.6)\gevc$ for 2011\,(2012) data and be 
inconsistent with originating from any PV. A multivariate algorithm~\cite{BBDT} is used for the identification of
secondary vertices consistent with the decay of a \bquark hadron. 

In this analysis, it is important to minimise the 
variation of the selection efficiency over the phase space of the decays of interest. Trigger signals are associated with reconstructed particles. Selection requirements can therefore be made on whether the decision was due to the signal candidate, other particles produced in the $pp$ collision or a combination of both. If it is required that the hardware trigger requirements are satisfied by a high-transverse-energy hadron belonging to the signal decay chain, a strong variation of the efficiency over the phase space is observed. Consequently, the strategy employed is that signal candidates are selected from events in which the hardware trigger requirements are satisfied by other activity in the event~\cite{LHCb-DP-2012-004}. In that case, the variation of the efficiency over the phase space is contained within 5\%.

The events passing the trigger requirements are then filtered in two stages. Initial requirements are applied to further 
reduce the size of the data sample before a multivariate selection is implemented. Selection requirements based on topological variables, such as the flight distance of the \bquark-baryon candidate, are used as the main discriminants. To reduce the variation of selection efficiency over the phase space of the decays of interest (a significant source of systematic uncertainty in the final result), only loose requirements are made on the transverse momenta of the daughter particles, $\pt >250\mevc$. 

%
%

The neutral \bquark-baryon candidates, henceforth denoted  $X_b$,  are formed from a proton candidate selected with 
particle identification (PID) requirements and three additional charged tracks. When more than one PV is reconstructed, 
the $X_b$ candidate is associated with the PV with which it forms the smallest $\chisqip$, where $\chisqip$ is the difference in $\chisq$ of a given PV reconstructed
with and without the considered $X_b$ candidate.  
Each of the four tracks of the final state is required to have $\ptot<100\gevc$, a value beyond which there is little 
pion/kaon/proton discrimination, and $\chisqip > 16$. 
The  $X_b$ candidates are then required to form a vertex with a fit quality $\chisqvtx < 20$ with 5 degrees of freedom and be significantly 
separated from any PV with $\chisqfd > 50$, where $\chisqfd$ is the square of the flight-distance significance. To remove backgrounds from higher-multiplicity decays, the difference in \chisqvtx when adding any other track must be greater than 4. The  $X_b$  candidates must have $\pt>1.5\gevc$ and invariant mass within the range 
$5340 < m(phhh) < 6400 \mevcc$, where $h$ stands for either a charged pion or kaon. They are further required to be consistent with originating from a PV, quantified by both the \chisqip and the ``pointing angle'' 
 between the reconstructed momentum of the $b$-hadron and the vector defined by the associated PV and the decay vertex. 
 Finally, PID requirements are applied to provide discrimination between kaons 
and pions in order to assign the candidates to one of the five different final-state spectra \ppipipi, \pKpipi, \pKKpi, \pKpiK and \pKKK. 

%
%

There are three main categories of background that contribute significantly in the selected invariant mass regions: the so-called signal ``cross-feed''  backgrounds resulting from a misidentification of one or more final-state particles; the charmless decays of neutral \B mesons to final states containing four charged mesons, where a  
pion or a kaon is misidentified as a proton; and the combinatorial backgrounds, which result from a random association of unrelated tracks. The pion and kaon PID requirements that define mutually exclusive samples are optimised to reduce the signal cross-feed background, and hence to maximize the observation of the signal. The charmless \B-meson decays are identified by 
reconstructing the invariant mass distributions of candidates reconstructed with a pion or kaon mass instead of the proton mass hypothesis, in the data high-mass sidebands, defined as $m_{\rm sideband} < m(\phhh) < 6400 \mevcc$, where $m_{\rm sideband}$ = 5680 \mevcc for \ppipipi, \pKKpi final states and $m_{\rm sideband}$= 5840 \mevcc for \pKpipi, \pKpiK, \pKKK final states. This background contribution is reduced by the optimisation of the proton PID requirement.

In order to reject combinatorial backgrounds, multivariate discriminants based on a boosted decision tree~(BDT)~\cite{Breiman} with the AdaBoost algorithm~\cite{AdaBoost} have been 
designed. Candidates from simulated \LbToppipipi  events and 
the data high-mass sideband are used as the signal and background training samples, respectively. This high-mass sideband region is chosen so that the sample is free of signal cross-feed background.  The samples are divided into two data-taking periods 
and  further subdivided into two equally sized subsamples.  Each subsample is then used to train an independent discriminant. In the subsequent
 analysis the BDT trained on one subsample is used to select candidates from the other subsample, in order to avoid bias. 
 
The BDTs have as input discriminating quantities the \pt, $\eta$, \chisqip, \chisqfd, 
pointing angle and \chisqvtx of the $X_b$ candidate; the smallest  change in the  \bquark-baryon \chisqvtx when adding any other track from the event; the sum of the \chisqip of the four tracks of the final state; and the  \pt asymmetry
\begin{equation}
p_{\rm T}^{\rm asym} = \frac{p_{\rm T}^\B - p_{\rm T}^{\rm cone}}{p_{\rm T}^\B + p_{\rm T}^{\rm cone}} \,,
\end{equation}
where $p_{\rm T}^{\rm cone}$ is the transverse component of the sum of all particle momenta inside a 1.5\rad cone in $\eta$ and $\phi$ space around the  \bquark-baryon candidate 
direction. The $p_{\rm T}^{\rm asym}$ of the signal candidates are preferentially distributed towards high values. The BDT output is determined to be uncorrelated with the position in the 
phase space of the decay of interest.  

The selection requirement placed on the output of the BDTs is independently optimised for the seven decays of interest by maximising the figure of 
merit~\cite{Punzi:2003bu}
\begin{equation}
\mbox{FoM} = \frac{\eps_{\rm sig}}{\frac{a}{2} + \sqrt{N_B }} \,,
\end{equation}

\noindent where the signal efficiency ($\eps_{\rm sig}$) is estimated from the simulation and $N_B$ represents the number of expected background 
events for a given selection, which is calculated by fitting the high-mass sideband of the data sample, and extrapolating the yield into the signal region defined as the invariant mass window covering $\pm$ 3 times the measured signal width.
The value $a=2$ is used in this analysis; it is found that varying this value up to 5 does not significantly change the result. A common 
optimisation of the BDT criteria is found, resulting in a signal efficiency of order 70\%.  

%
%
A number of background contributions consisting of fully reconstructed \bquark-baryon decays into the two-body $\Lc \had$, $\Xicp \had$, three-body $\D p \had$ or  $(\cquark\cquarkbar) p \had$ 
combinations, where $(\cquark\cquarkbar)$ represents a charmonium resonance, may produce the same final state as the signal. Hence, they will have the same \bquark-baryon candidate invariant mass distribution as the signal candidates, as well as a similar selection efficiency.
The presence of a misidentified hadron in the \D , \Lc and \Xicp decay also produces peaking background under the signal.  Therefore, the following decay channels are explicitly reconstructed
under the relevant particle hypotheses and vetoed by means of a requirement on the resulting invariant mass, in all experimental spectra: \Lc ($\to pK^-\pi^+, p\pi^+\pi^-, pK^+K^-$), \Xicp ($ \to pK^-\pi^+$), \Dp ($\to K^-\pi^+\pi^+$), \Dsp ($\to K^-K^+\pi^+$), \Dz ($\to K^{\mp}\pi^{\pm},\pi^+\pi^-, K^+K^-$), \chiczero and \jpsi ($\to \pi^+\pi^-, K^+K^-$).

%
%

The same set of trigger, \pid and BDT requirements is applied to the normalisation mode \LbToLcpiLcTopKpi to cancel out most of 
the systematics effects related to the selection criteria. Candidates whose $pK^-\pi^+$ invariant mass is in the range $2213 < m(pK^-\pi^+) <  2313$ \mevcc 
are retained as normalisation-mode candidates.  Conversely, events outside this interval belong to the signal \pKpipi spectrum, 
again ensuring statistically independent samples for the simultaneous fit. 

%
%

The fraction of events containing more than one candidate is below the percent level.  The candidate to be retained in each event is chosen randomly and reproducibly.

\section{Simultaneous fit}
\label{sec:fit}

A simultaneous unbinned extended maximum likelihood fit is performed to the $\bquark$-hadron candidate invariant mass distributions under each 
of the five sets of mass hypotheses for the final-state tracks and the normalisation channel candidates. The data samples are further split according to the year of data taking. 
The components of the model include, in addition to the signal decays,  the partially reconstructed five-body \Xb decays, the signal and background 
cross-feeds, the four- and the five-body decays of \B-mesons and the combinatorial background. The independent data samples constructed for each experimental reconstructed spectrum are fitted simultaneously. For each sample, the likelihood is expressed as
\begin{equation}
 \displaystyle \ln{\cal L}  =  \sum_{i} \ln \left( {\sum_{j} N_j P_{j,i}} \right) -\sum_{j}N_j  
\end{equation}
\noindent where $N_{j}$ is the number of events related to the component $j$ and $P_{i}$ the probability of the candidate $i$.

%
%

\subsection{Fit model} 
\label{sec:model}

The signal decays are modelled as the sum of two Crystal Ball (CB) functions~\cite{Skwarnicki:1986xj}.
These two CB functions share peak positions and widths but have independent power-law tails on opposite sides of the peak.  
The \Lb mass parameter, corresponding to the most probable value of the double-CB function, is free in the fit and is shared among all invariant mass spectra. The difference between the \Xibz and \Lb masses is also a 
shared parameter and is constrained to the measured value in Ref.~\cite{PDG2017}.  

The ratio of the experimental widths of the signal decay functions is constrained using Gaussian prior probability distributions included in the likelihood, with parameters
obtained from the fit to simulated events. 
The measured \LbTopKpipi width in the 2012 data-taking sample is chosen as the reference (measured to be $\sigma = 16.47 \pm 0.22$ \mevcc). The other parameters of the CB components are obtained by a simultaneous 
fit to simulated samples, and are fixed to those values in the nominal fits to the data. 


The cross-feed backgrounds are modelled by the sum of two CB functions, the parameters of which are determined from simulated samples. 
All cases resulting from the misidentification of either one or two of the final-state particles are considered. The relative yield
 of each misidentified decay is constrained with respect to the yield of the corresponding correctly identified decay and the known misidentification probabilities. 
The constraints are implemented using Gaussian prior probability distributions included in the likelihood. Their mean values 
are obtained from the ratio of selection efficiencies and their widths include uncertainties originating from the finite size of the 
simulated events samples as well as the systematic uncertainties related to the determination of the PID efficiencies.


The backgrounds resulting from four- or five-body decays of \B mesons are identified in each spectrum by a dedicated fit to the candidates 
in the high-mass sideband, reconstructed under the hypothesis of a kaon mass for the proton candidates. The relative yield of each 
decay is then constrained in the simultaneous fit from its observed abundance in the high-mass sidebands. The invariant mass 
distributions are modelled by the sum of two CB functions, the parameters of which are determined 
from simulated events.  


Partially reconstructed backgrounds where a neutral pion is not reconstructed, such as \Lb, \Xibz \to \phhhpiz, are modelled by means of generalised ARGUS functions~\cite{Albrecht:1990cs} convolved with a Gaussian resolution function. The Gaussian width is taken as the signal \LbTopKpipi width parameter. The parameters of the ARGUS function are shared among all invariant mass spectra and are determined directly from the fit, except for the threshold, which is given by $m(X_b) - m(\pi^0)$.  Radiative decays such as \LbToppietap and  \LbTopKetap (\EtapTopipig) are modelled separately using the same functional form but where the parameters are determined using simulated events. The decay modes \LbTopKpipipiz where a pion is misidentified as a kaon can significantly contribute to the \pKKpi and \pKpiK spectra. They are modelled with an empirical (histogrammed) function determined from the partially reconstructed background candidates in the normalisation channel. 

Finally, the combinatorial background is modelled by a linear function whose slope is shared among the invariant mass spectra. An exponential function is used as an alternative model in order to estimate any systematic effect related to this choice of modelling. 

\subsection{Fit results} 
\label{sec:fitresults}

\begin{figure}[!t]
  \centering
  \includegraphics[width=\columnwidth]{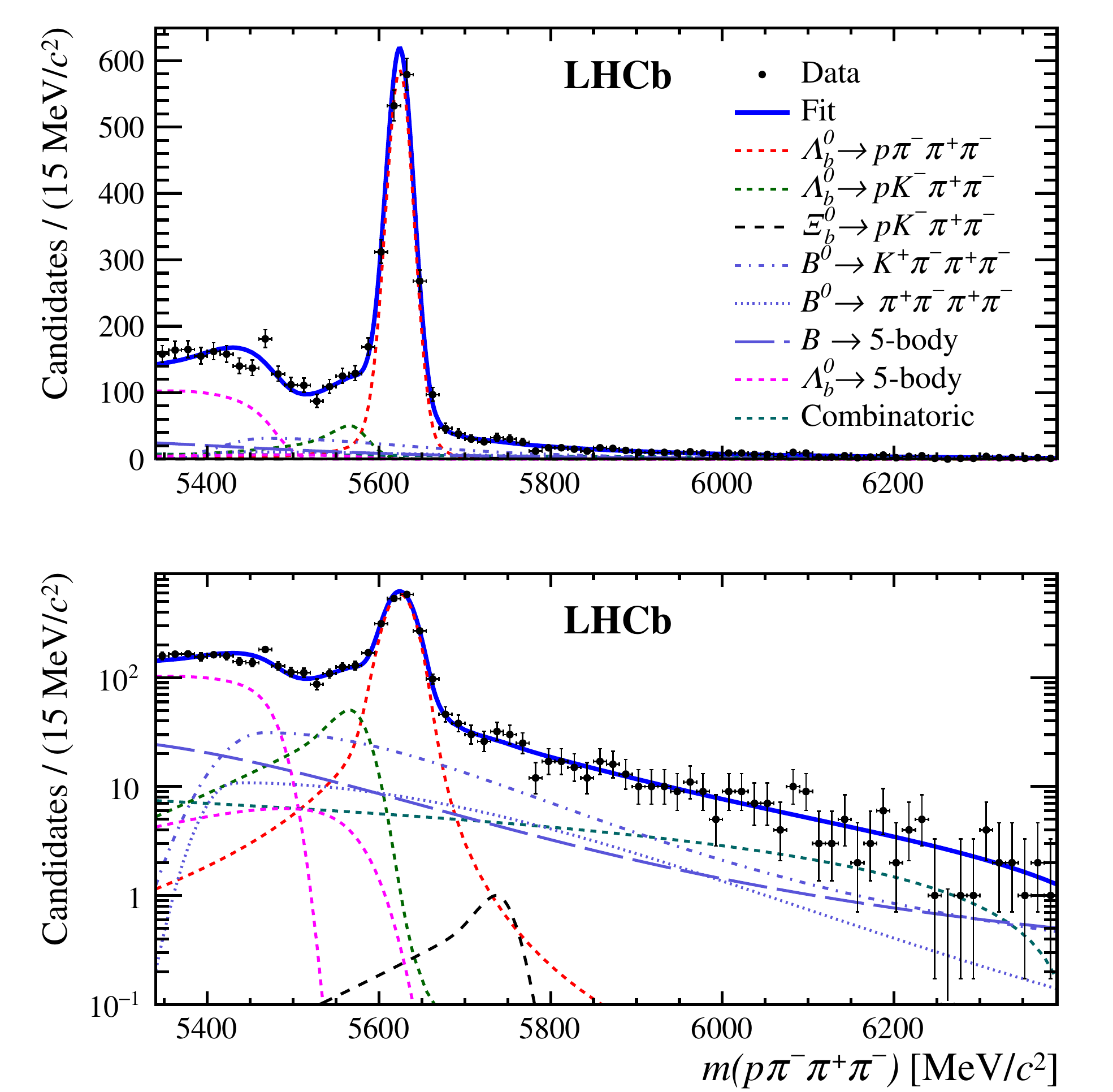}
  \caption{Results of the fit to the \ppipipi candidate mass spectrum with (top) linear and (bottom) logarithmic scales. The different components employed in the fit are indicated in the legend. The $\Lb\to$ 5-body legend describes two components, the radiative partially reconstructed background \LbToppietap and the partially reconstructed background \LbToppipipipiz where a $\pi^0$ is not reconstructed. The latter has a lower-mass endpoint.}
  \label{fitresults_all_pipipi}
\end{figure}

\begin{figure}[!t]
  \centering
  \includegraphics[width=\columnwidth]{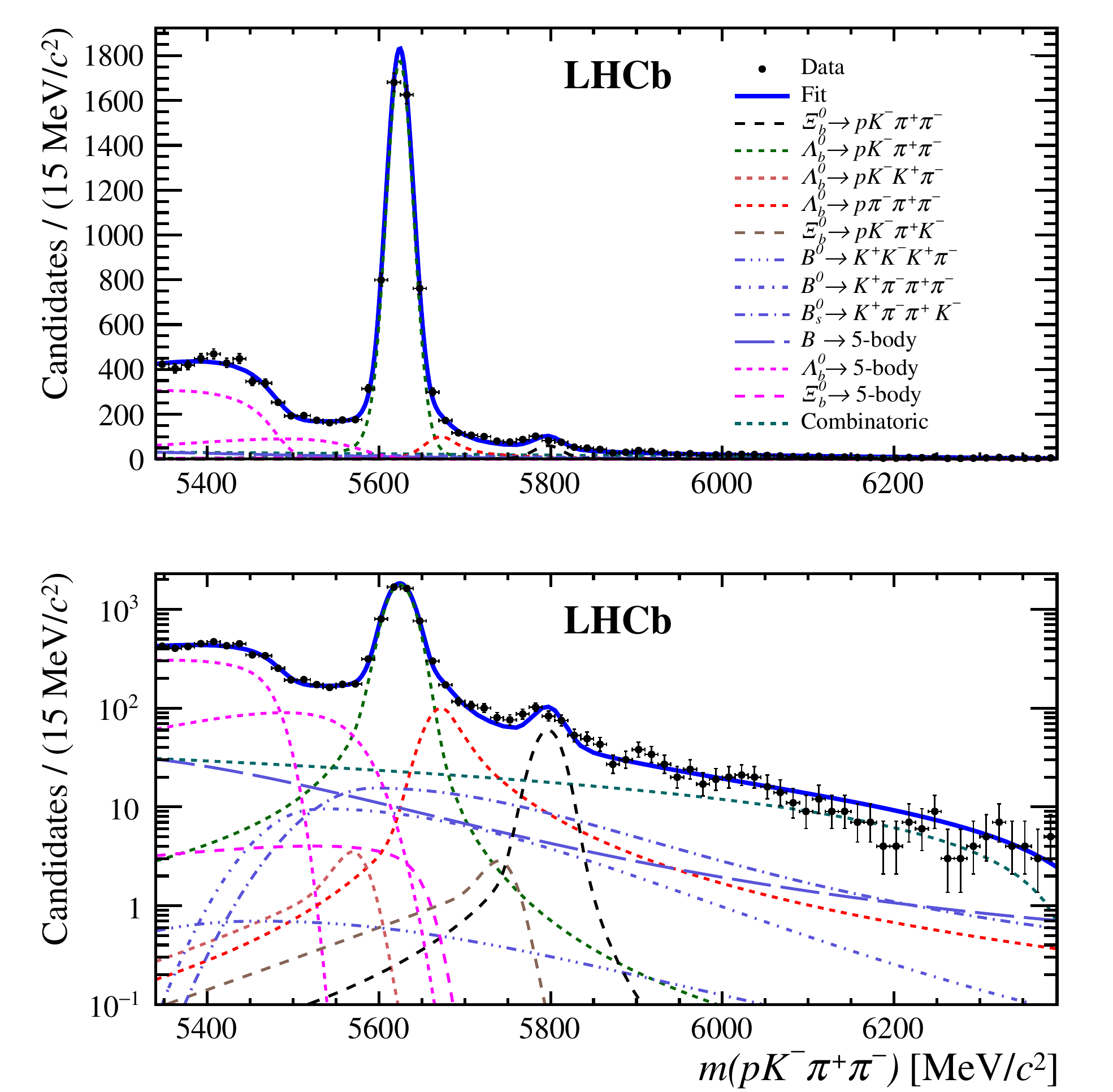}
  \caption{Results of the fit to the \pKpipi candidate mass spectrum with (top) linear and (bottom) logarithmic scales. The different components employed in the fit are indicated in the legend. The $\Lb\to$ 5-body legend describes two components,  the radiative partially reconstructed background  \LbTopKetap and the partially reconstructed background \LbTopKpipipiz where a $\pi^0$ is not reconstructed. The latter has a lower-mass endpoint.}
  \label{fitresults_all_Kpipi}
\end{figure}

\begin{figure}[!t]
  \centering
  \includegraphics[width=\columnwidth]{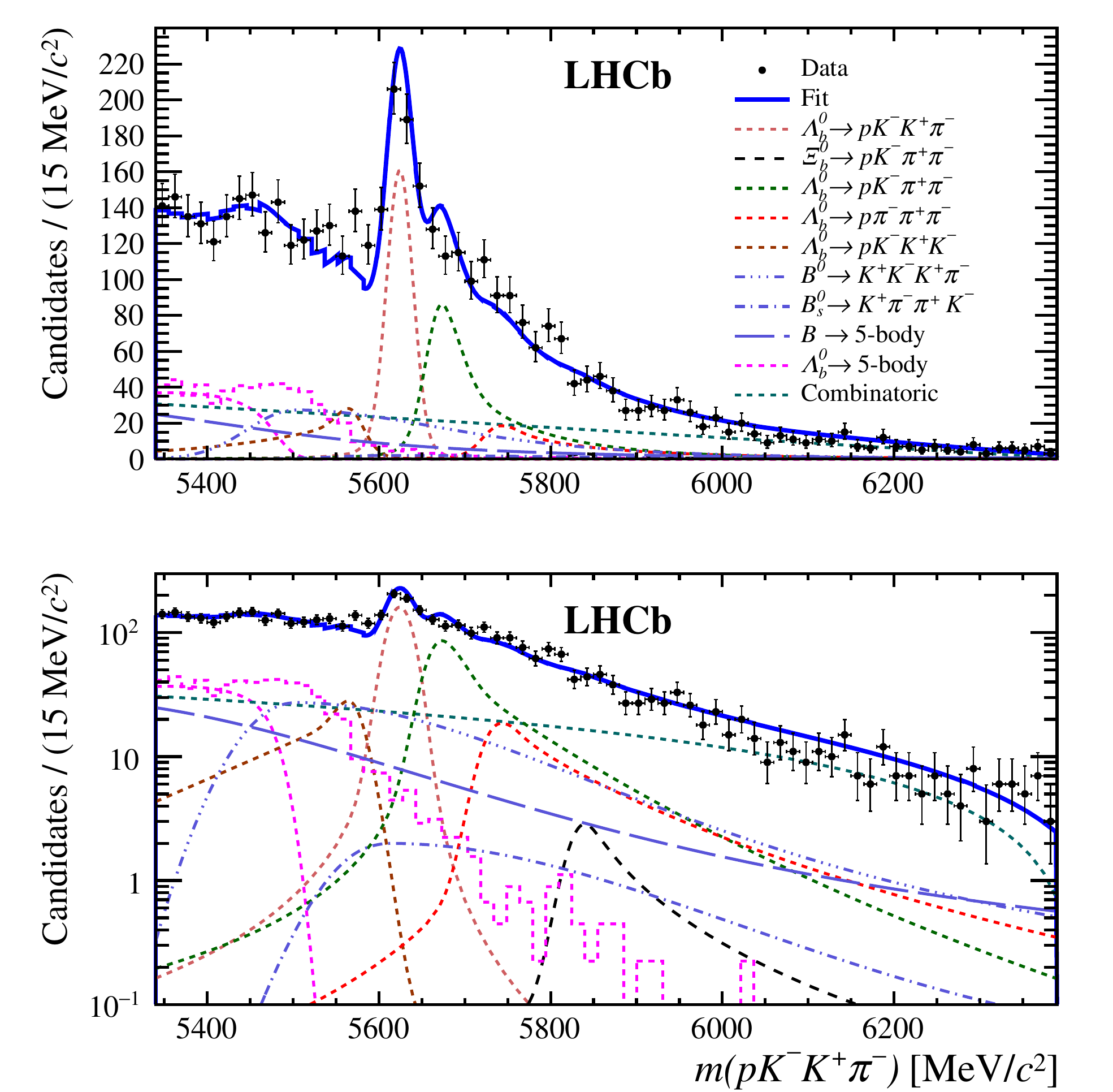}
  \caption{Results of the fit to the \pKKpi candidate mass spectrum with (top) linear and (bottom) logarithmic scales. The different components employed in the fit are indicated in the legend. The $\Lb\to$ 5-body legend describes two components where a $\pi^0$ is not reconstructed, the partially reconstructed background \LbTopKpipipiz where a pion is misidentified as a kaon and the partially reconstructed background \LbTopKKpipiz.}
  \label{fitresults_all_KKpi}
\end{figure}

\begin{figure}[!t]
  \centering
  \includegraphics[width=\columnwidth]{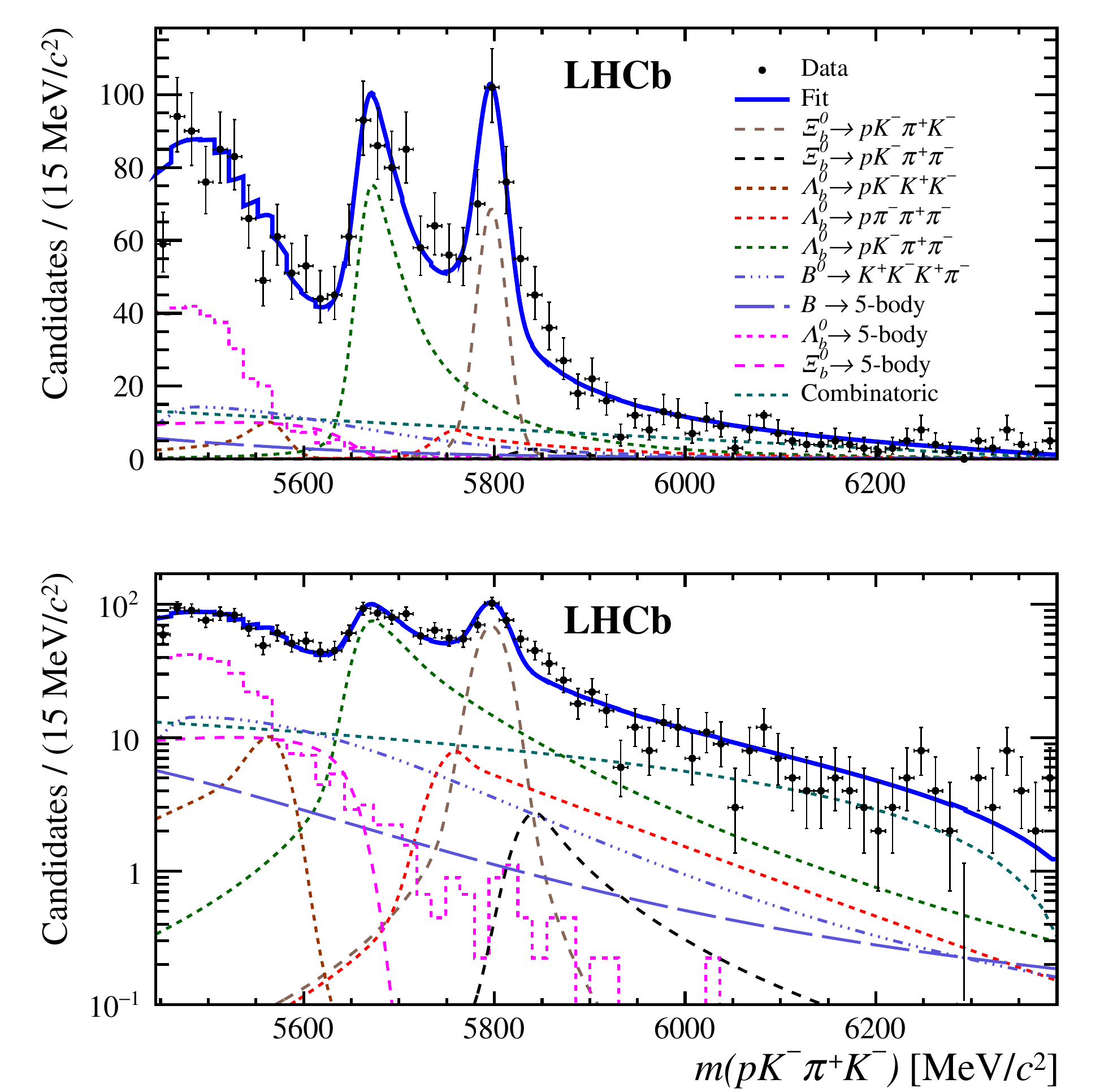}
  \caption{Results of the fit to the \pKpiK candidate mass spectrum with (top) linear and (bottom) logarithmic scales. The different components employed in the fit are indicated in the legend.}
  \label{fitresults_all_KpiK}
\end{figure}

\begin{figure}[!t]
  \centering
  \includegraphics[width=\columnwidth]{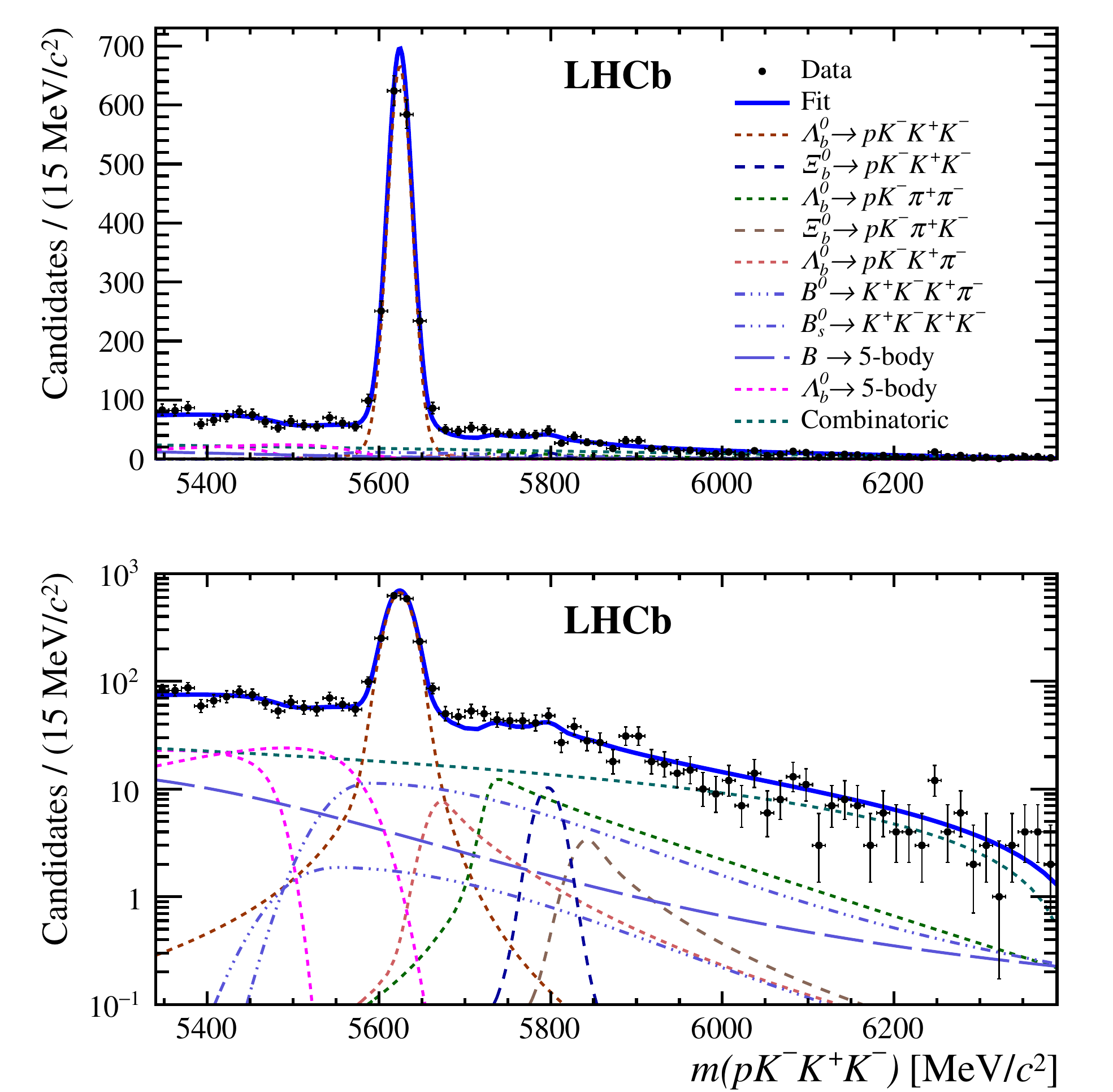}
  \caption{Results of the fit to the \pKKK candidate mass spectrum with (top) linear and (bottom) logarithmic scales. The different components employed in the fit are indicated in the legend. The $\Lb\to$ 5-body legend includes two decays, partially reconstructed \LbTopKKKg and \LbTopKKKpiz, where the \g and $\pi^0$ are not reconstructed.}
  \label{fitresults_all_KKK}
\end{figure}

\begin{figure}[!t]
  \centering
  \includegraphics[width=\columnwidth]{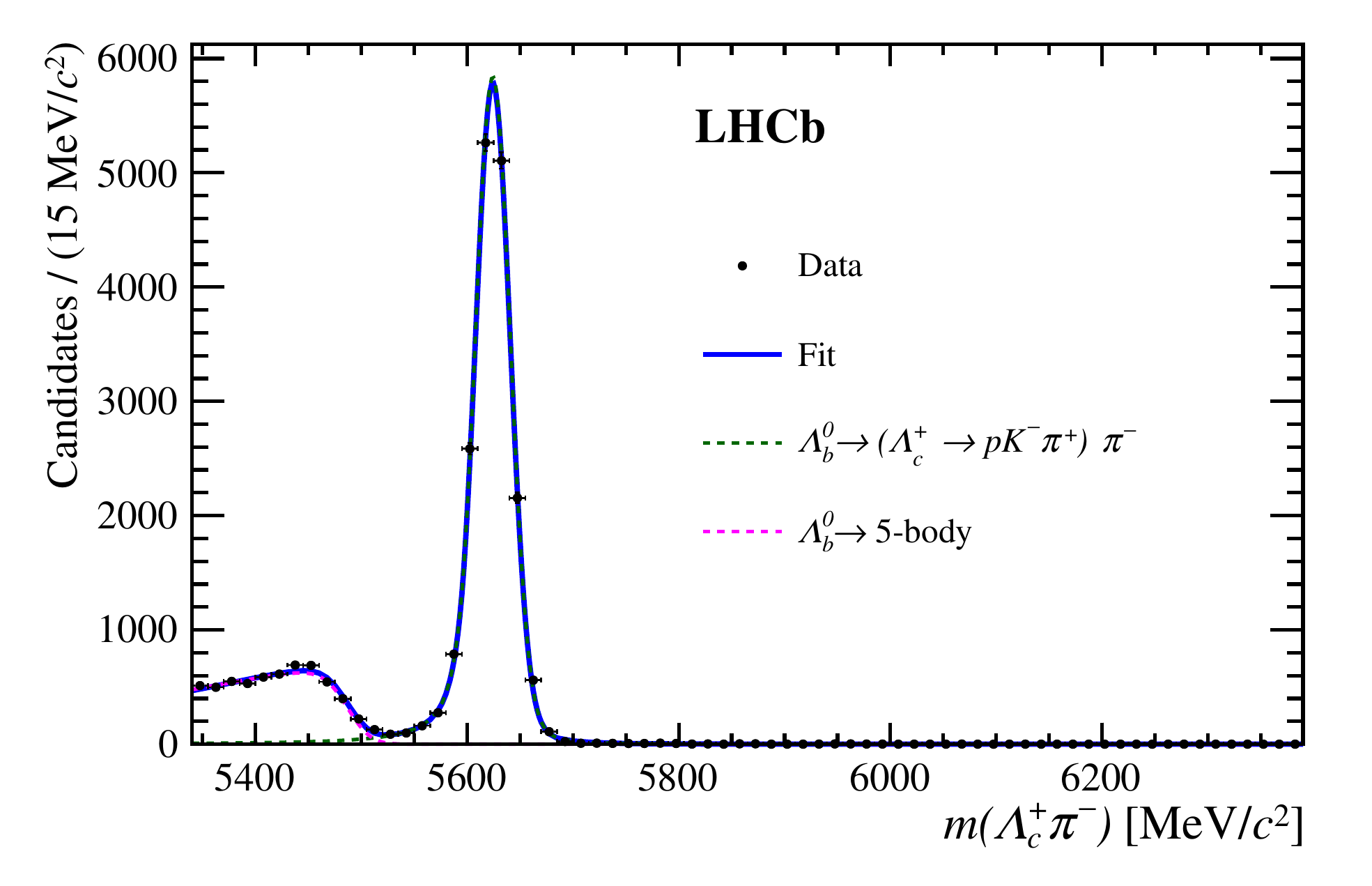}
  \caption{Results of the fit to the $\Lc\pi^-$ candidate mass spectrum on linear scale. The different components employed in the fit are indicated in the legend.}
  \label{fitresults_all_cc}
\end{figure}

Figures~\ref{fitresults_all_pipipi} to \ref{fitresults_all_KKK} display the 
fit results of the simultaneous fit to the invariant mass spectra of the five final states using the whole data sample. Figure~\ref{fitresults_all_cc} displays the result of the fit to the normalisation channel. 
The signal yields for each decay channel are shown in Table~\ref{tabfitresults}. The fit model provides an overall satisfactory description of the data. However, differences between the data and the fit model 
can be noted in the high-mass sidebands of Figs. \ref{fitresults_all_Kpipi}, \ref{fitresults_all_KpiK} and \ref{fitresults_all_KKK}. The significance of the disagreement is not larger than two standard deviations. Those discrepancies are covered within the size of the variations considered in the evaluation of the systematic uncertainties. 

\begin{table}[!t]
  \renewcommand{\arraystretch}{1.1}
  \caption{Signal yields for each decay mode, determined by summing the fitted yields in each year of data taking. The signal ($S$) to background ($B$, adding all sources) ratios 
  in an invariant mass window, covering $\pm$ 3 times the measured signal widths, are provided. The corresponding invariant mass ranges are reported in the fourth column.}
  \label{tabfitresults}
  \begin{center}
    \begin{tabular}{l D{p}{\,\pm\,}{3.3} D{p}{\,\pm\,}{3.3} cccc}
    \hline
   Decay mode             &  \multicolumn{1}{c}{Signal yield}  & \multicolumn{1}{c}{$S/B$}      &  $\pm 3\sigma$ range (\mevcc) \\
   \hline 
   \LbToppipipi           & 1809 p \textcolor{white}{0}48   & 4.9\textcolor{white}{0}  p 0.3    & [5573.9, 5674.6] \\
   \LbTopKpipi            & 5193 p \textcolor{white}{0}76   & 7.7\textcolor{white}{0}  p 0.4    & [5574.4, 5674.2]  \\
   \LbTopKKpi             & 444  p \textcolor{white}{0}30   & 0.71 p 0.06                       & [5577.4, 5671.1] \\  
   \LbTopKKK              & 1706 p \textcolor{white}{0}46   & 8.1\textcolor{white}{0}  p 0.7    & [5579.0, 5674.6]  \\
   \XibzTopKpipi          & 183  p \textcolor{white}{0}22   & 0.59 p 0.09                       & [5747.9, 5846.2] \\
   \XibzTopKpiK           & 199  p \textcolor{white}{0}21   & 0.81  p 0.10                      & [5747.4, 5846.2]  \\
   \XibzTopKKK            & 27   p \textcolor{white}{0}14   & 0.14 p 0.08                       & [5752.7, 5840.8] \\
   \LbToLcpiLcTopKpi      & 16518 p 133                     &  \multicolumn{1}{c}{---}          & [5573.7, 5674.8] \\
    \hline 
    \end{tabular}
    \end{center}
\end{table}

All signals that were searched for are established unambiguously with the exception 
of the \XibzTopKKK decay. The signal-to-background ratios vary from mode to mode following the hierarchy of the branching fractions and are summarized in Table~\ref{tabfitresults}.


\section{Determination of the signal efficiencies}
\label{sec:efficiency}

The experimentally determined result for each four-body signal decay is the quantity $R$, defined as 

\begin{eqnarray}
\label{masterformula}
R(\XbzTophhhz) & \equiv &\frac{ \BF (\XbzTophhhz) }{ \BF (\LbToLcpi) } \cdot \frac{f_{X_b}}{f_{\Lb}}, \nonumber \\ \\
              &    =   & \frac{ \epsilon_{\LbToLcpi}^{\mbox{\small{geo.} }} }{ \epsilon_{\XbzTophhhz}^{\mbox{\small{geo.}}} } \cdot \frac{ \epsilon_{\LbToLcpi}^{\mbox{\small{sel.} }} }{ \epsilon_{\XbzTophhhz}^{\mbox{\small{sel.}}} } \cdot \frac{ \epsilon_{\LbToLcpi}^{\mbox{\scriptsize{PID} }} }{ \epsilon_{\XbzTophhhz}^{\mbox{\scriptsize{PID}}} } \cdot \frac{ 1 }{ \epsilon_{\XbzTophhhz}^{\mbox{\scriptsize{veto}}} } \cdot \frac{ \Y_{\XbzTophhhz} }{ \Y_{\LbToLcpi} }, \nonumber
\end{eqnarray}

\noindent where \BF represents the relevant branching fraction and $f_{X_b}/{f_{\Lb}}$ is the relative hadronisation fraction of $\bquark \to X_b$ with respect to $\bquark \to \Lb$.
From left to right, the ratios of efficiencies are related to the geometrical acceptance, the selection criteria, the PID requirements and the veto of charm and charmonium backgrounds. The measured signal and normalisation channel yields are represented by $\Y_{\XbzTophhhz}$ and $\Y_{\LbToLcpi}$.

\par 

The efficiencies are determined from simulated signal events that have been generated with an arbitrary mixture of phase-space decays and 
quasi-two-body amplitudes, which feature the production of intermediate resonances close to their kinematic threshold. For instance, the \LbTopKpipi decay proceeds 
in the simulation of quasi-two-body amplitudes via the decays \ensuremath{\Lb \to \PLambda^{*}(1520)^0 \rho(770)^0}, \ensuremath{\Lb \to \PLambda^{*}(1520)^0 f_2(1270)} or $\Lb \to N^{*}(1520)^0 K^{*}(892)$.
In principle,  the selection efficiency of each decay mode depends on the phase-space coordinates, but the actual dynamics of the decays is {\it a priori} 
unknown and a data-driven correction of the efficiency determination with simulated events would be required as was done in Ref.~\cite{LHCb-PAPER-2017-010}. 
However, the candidate selection has been designed without relying on the kinematics of the daughter particles in the decay. The candidates selected such that the hardware trigger 
is satisfied independently of the signal particles, provide a sample with an efficiency that is, to a very good approximation, constant over the phase space of the decays.
The residual variation of the efficiency over the phase space is consequently addressed as a systematic 
uncertainty.  

\par 

The imperfections of the simulation are corrected for in several respects. Inaccuracies of the tracking simulation and the PID simulation are 
 mitigated by a weighting of the simulation to match the efficiencies measured in the data calibration 
 samples~\cite{LHCb-DP-2012-003}. 
 The uncertainties related to these corrections are propagated to the branching fraction measurements as systematic uncertainties. Other inaccuracies in the 
  simulation are addressed as systematic uncertainties and discussed in Section~\ref{sec:systematics}.  
  A number of two- or three-body invariant mass criteria have been used to veto charm and charmonium resonances. The efficiency of these 
  vetoes is determined {\it a posteriori} on the data samples by inferring the number of signal candidates vetoed by each mass criterion from a linear interpolation
of the invariant mass distribution reconstructed under the relevant mass hypotheses of the final-state particles.

  \par 
  
  Table~\ref{FinalEff} shows the ratios of efficiencies for the 2011 and 2012 data-taking periods, 
  necessary to derive the branching fraction values relative to the normalisation channel \LbToLcpi . The associated uncertainties are 
  propagated as systematic uncertainties in the derivation of the branching fractions.

\begin{table}[!t]
  \setlength{\tabcolsep}{7pt}
  \caption{Ratios of the normalisation decay mode efficiencies, relative to the signal decay mode as used in Eq. \eqref{masterformula}, for (first row) 2011 and (second row) 2012. The last column shows the efficiency of the veto of charm and charmonium backgrounds (applied to the signal mode only), as discussed in the text. Since the \XibzTopKKK decay mode is not observed, the veto efficiency is determined with the simulated data sample. 
The difference between the simulation value and the average veto efficiency measured on other \Xibz modes is reported in the table as the uncertainty.}
  \label{FinalEff}
  \begin{tabular*}{\textwidth}{lcccc}
    \hline
    Decay mode     &     \multicolumn{4}{c}{Ratios of efficiencies}                \\
                   &     Acceptance & Selection & \pid          & Vetoes \\
    \hline
    \multirow{2}{*}{\LbToppipipi}  &     1.070 $\pm$ 0.003 & 0.433 $\pm$ 0.011 & 1.018 $\pm$ 0.013 & 0.693 $\pm$ 0.028   \\
                                   &     1.050 $\pm$ 0.004 & 0.425 $\pm$ 0.009 & 1.046 $\pm$ 0.010 & 0.712 $\pm$ 0.017  \\
    \hline
    \multirow{2}{*}{\LbTopKpipi}   &     1.020 $\pm$ 0.003 & 0.438 $\pm$ 0.011 & 0.922 $\pm$ 0.012 & 0.758 $\pm$ 0.032  \\
                                   &     1.004 $\pm$ 0.004 & 0.432 $\pm$ 0.009 & 0.958 $\pm$ 0.009 & 0.744 $\pm$ 0.016  \\
    \hline
    \multirow{2}{*}{\LbTopKKpi}    &     0.978 $\pm$ 0.003 & 0.462 $\pm$ 0.012 & 0.846 $\pm$ 0.011 & 0.742 $\pm$ 0.099  \\
                                   &     0.970 $\pm$ 0.004 & 0.468 $\pm$ 0.010 & 0.874 $\pm$ 0.008 & 0.765 $\pm$ 0.050  \\
    \hline
    \multirow{2}{*}{\LbTopKKK}     &     0.928 $\pm$ 0.003 & 0.445 $\pm$ 0.012 & 0.783 $\pm$ 0.010 & 0.751 $\pm$ 0.036  \\
                                   &     0.916 $\pm$ 0.003 & 0.452 $\pm$ 0.010 & 0.801 $\pm$ 0.007 & 0.787 $\pm$ 0.026  \\
    \hline
    \multirow{2}{*}{\XibzTopKpipi} &     1.019 $\pm$ 0.003 & 0.431 $\pm$ 0.011 & 0.902 $\pm$ 0.011 & 0.652 $\pm$ 0.082  \\
                                   &     1.009 $\pm$ 0.004 & 0.424 $\pm$ 0.009 & 0.917 $\pm$ 0.008 & 0.659 $\pm$ 0.109  \\
    \hline
    \multirow{2}{*}{\XibzTopKpiK}  &     0.979 $\pm$ 0.003 & 0.434 $\pm$ 0.011 & 0.829 $\pm$ 0.010 & 0.689 $\pm$ 0.074  \\
                                   &     0.969 $\pm$ 0.004 & 0.450 $\pm$ 0.010 & 0.847 $\pm$ 0.008 & 0.752 $\pm$ 0.081  \\
    \hline
    \multirow{2}{*}{\XibzTopKKK}   &     0.929 $\pm$ 0.003 & 0.425 $\pm$ 0.011 & 0.764 $\pm$ 0.009 & \multirow{2}{*}{0.819 $\pm$ 0.123}  \\ 
                                   &     0.922 $\pm$ 0.003 & 0.429 $\pm$ 0.009 & 0.771 $\pm$ 0.007 & \\
    \hline

\end{tabular*}
\end{table}


\section{Systematic uncertainties}
\label{sec:systematics}

The systematic uncertainties are largely reduced by normalising the branching fraction 
measurements with respect to that of the decay channel \LbToLcpiLcTopKpi. The remaining 
sources of systematic uncertainties and the methods used to estimate them 
are described in this section. Tables \ref{tab:yieldseff2011} and \ref{tab:yieldseff2012} provide the yields measured by the 
fit, the related statistical uncertainties, the overall efficiency, as well as the systematic uncertainty for each decay, for 2011 and 2012 data, respectively. 
The other sources of systematic uncertainty, which are not reported here, have negligible impact on the measurements.

\subsection{Fit model uncertainties}
\label{sec:fitmodelsyst}

Uncertainties related to the fit model result from uncertainties in the values of the parameters taken 
from the simulation as well as from the choice of the functional forms used to describe the various 
components of the model.

\par 

The systematic uncertainties related to the parameters fixed to values determined from simulated 
events are obtained by repeating the fit with the parameters allowed to vary according to their 
uncertainties using pseudoexperiments. The fixed parameters that are driving the shape of the tails of the functional forms describing 
signal channels, cross-feeds and \B backgrounds distributions are estimated from a simultaneous fit of the simulated 
events of these categories. The parameters are then varied according to the covariance matrix 
obtained from simulated events. The nominal fit is then performed on this ensemble of pseudoexperiments 
and the distribution of the difference between the yield determined in each of these fits and that 
of the nominal fit is in turn fitted with a Gaussian function. The systematic uncertainty associated 
with the choice of the value of each signal parameter from simulated events is then assigned as the linear 
sum of the absolute value of the mean of the Gaussian and its width. The variation of the fixed parameters 
of a functional form covers any reasonable variation of that shape. 

\par

The combinatorial background is modelled by a linear function.  This model is substituted by an exponential 
form in the fit to the data. Pseudoexperiments based on the latter model are fit with the nominal model.  
The value of the uncertainty is computed as the linear sum of the mean of the resulting distribution and its RMS. 

\par 

The mixture of quasi-two-body and phase-space decays that has been 
used to generate the simulation samples is a source of systematic uncertainty. The true signal 
dynamics ({\it a priori} unknown) lies between two extreme cases: the decays are saturated by 
quasi-two-body amplitudes or are fully described by a uniform amplitude over phase space. The shapes used to model all signal modes and cross-feeds are weighted 
according to these two extreme cases and the range of variation of the fit results obtained under the 
two conditions is taken as the corresponding systematic uncertainty estimate. In addition, the data-driven 
kinematics-dependent \pid corrections, applied to the \pid efficiencies, obtained in the simulation to 
match the data, are also used to weight the functional forms of all the components of 
the fit model derived from simulated events.

\par

The total systematic uncertainty of the fit model is given by the sum in quadrature of all the contributions. It 
is mostly dominated by the shape parameters fixed to values determined from simulated events.

\subsection{Selection efficiency uncertainties}
\label{sec:effsyst}

The most significant source of systematic uncertainty is related to the control of the variation of the candidate selection efficiency over the phase space 
of the decays of interest. The systematic uncertainties coming from the determination of the efficiencies are larger than the statistical uncertainties for a few modes. 
Their estimation relies on the simulation of the two extreme dynamics of each decay, namely intermediate resonances close to the kinematic threshold (\eg $\PLambda^{*}(1520)^0 \rho(770)^0$,  $\PLambda^{*}(1520)^0 f_2(1270)$ or $N^{*}(1520)^0 K^{*}(892)$ for 
\LbTopKpipi simulated signal events) or uniformly populated phase-space decays. The difference in efficiency measured between these two cases is 
examined for all elements of the signal candidate selection procedure: geometrical acceptance, reconstruction and selection, 
trigger, \pid and  BDT criteria. The individual ranges of variation are summed in quadrature to provide 
the total systematic uncertainty estimate, which is found to be the dominant source for most of the modes. The correlation between the
determinations for 2011 and 2012 data samples is taken into account in the combined measurement.

\par

The training of the BDT relies on simulated signal events. Potential 
inaccuracies in the simulation of the variables used in the BDT produce suboptimal 
discrimination of the multivariate tool. In addition, the $b$-hadron kinematics is a known source of differences between 
simulated events and data, and can further induce a bias in the signal efficiency determination. The systematic 
uncertainty due to this effect is estimated by weighting the simulated distributions of the \pt 
and $\eta$ of the $X_b$ candidates to match the distributions of the selected data for the normalisation channel.  The 
observed differences with the nominal selection efficiency are taken as the uncertainty estimates.

\par 

Uncertainties related to the efficiencies of the charm and charmonium resonance vetoes (discussed  in Section~\ref{sec:efficiency}) are dominated by the statistical uncertainties on the counting of the candidates in the two- or three-body invariant mass distributions before and after the veto criteria. It is analytically propagated to the branching fraction measurements and is a major source in the systematic uncertainty budget.

\begin{table}[!tb]
  \renewcommand{\arraystretch}{1.1}
  \caption{Yields and efficiencies of each signal decay with the statistical uncertainty, and systematic uncertainties related to the fit model and the efficiency determination, for the 2011 data samples.}
  \label{tab:yieldseff2011}
\newcolumntype{d}[1]{D{.}{.}{#1}}                                                                                                                                                                 
\newcolumntype{p}[1]{D{p}{\pm}{#1}}                                                                                                                                                                 

  \begin{tabular*}{\columnwidth}{l d{4.0} c p{0.2} p{0.2} p{0.2} }
    \hline
   Decay mode                &     \multicolumn{1}{c}{Yield} & Eff.\small{($10^{-3}$)} & \multicolumn{1}{c}{Stat.\small{(\%)}} & \multicolumn{1}{c}{Fit Model\small{(\%)}} & \multicolumn{1}{c}{Eff. Syst.\small{(\%)}} \\
    \hline

    \LbToppipipi         &     533   & 0.51 & p 4.8  & p 1.4  & p 5.2\\ 
    \LbTopKpipi          &     1679  & 0.64 & p 2.6  & p 1.1  & p 5.5\\ 
    \LbTopKKpi           &     120   & 0.68 & p 14   & p 8.5  & p 14\\
    \LbTopKKK            &     565   & 0.81 & p 4.7  & p 1.8  & p 6.4\\
    \XibzTopKpipi        &     65    & 0.57 & p 19   & p 3.5  & p 14\\
    \XibzTopKpiK         &     68    & 0.68 & p 17   & p 5.2  & p 12\\
    \XibzTopKKK          &     9     & 0.95 & p 83   & p 12.8 & p 16\\
    \hline
    \LbToLcpiLcTopKpi    &     5427  & 0.35 & p 1.4  & p 0.8  & \multicolumn{1}{c}{---}\\
    \hline
\end{tabular*}
\end{table}

\begin{table}[!tb]
  \renewcommand{\arraystretch}{1.1}

\newcolumntype{d}[1]{D{.}{.}{#1}}                                                                                                                                                                 
\newcolumntype{p}[1]{D{p}{\pm}{#1}}                                                                                                                                                                 
  \caption{Yields and efficiencies of each signal decay with the statistical uncertainty, and systematic uncertainties related to the fit model and the efficiency determination, for the 2012 data samples.}
   \label{tab:yieldseff2012}
  \begin{tabular*}{\columnwidth}{l d{5.0} c p{0.2} p{0.2}  p{0.2} }
    \hline
   Decay mode                &     \multicolumn{1}{l}{Yield} & Eff.\small{($10^{-3}$)} & \multicolumn{1}{l}{Stat.\small{(\%)}} & \multicolumn{1}{l}{Fit Model\small{(\%)}}  & \multicolumn{1}{l}{Eff. Syst.\small{(\%)}} \\

    \hline

    \LbToppipipi         &    1277  & 0.45 & p 3.2 & p 1.2  & p 4.8\\
    \LbTopKpipi          &    3515  & 0.53 & p 1.9 & p 1.3  & p 3.7\\
    \LbTopKKpi           &    324   & 0.57 & p 7.9 & p 5.9  & p 7.3\\
    \LbTopKKK            &    1141  & 0.70 & p 3.3 & p 1.4  & p 5.1\\
    \XibzTopKpipi        &    118   & 0.49 & p 16  & p 3.1  & p 18\\
    \XibzTopKpiK         &    131   & 0.60 & p 13  & p 5.8  & p 13\\
    \XibzTopKKK          &    19    & 0.79 & p 60  & p 10   & p 16\\
    \hline
    \LbToLcpiLcTopKpi    &  12226   & 0.29 & p 1.0 & p 0.8  & \multicolumn{1}{c}{---}\\
    \hline
\end{tabular*}
\end{table}
\vspace{1.5cm}

\section{Branching fraction measurements and concluding remarks}
\label{sec:conclusions}

Six decays are unambiguously observed. The  \XibzTopKKK decay mode is measured with a significance of $2.3\sigma$. Tables \ref{tab:bf2011} and \ref{tab:bf2012} summarise the relative branching fraction measurements determined from Eq.~\eqref{masterformula}, separately for the 2011 and 2012 data samples. The consistency of the two determinations of each decay mode for each year is quantified as the ratio of the signed difference of the central values over the quadratic sum of the related uncertainties. The two measurements are in fair agreement, namely better that $2.1$ statistical standard deviations in all cases.

\begin{table}[t!]
\setlength{\tabcolsep}{4.pt}
\renewcommand{\arraystretch}{1.1}
\caption{Measurements of the $R$ ratio from the (first row) 2011 and the (second row) 2012 data samples for \Lb decay modes expressed in percent as well as their combination. The three uncertainties are statistical, systematic related to the fit model and systematic related to the efficiency, respectively. The consistency of the two determinations for each year, denoted $\Delta$, is quantified as the ratio of the signed difference of the central values over the quadratic sum of the related uncertainties.}
\label{tab:bf2011}
\newcolumntype{p}[1]{D{p}{\,\pm\,}{#1}}

 \begin{tabular*}{\textwidth}{l c c c }
 \hline
$R$ (per decay)                 &  Value ($\%$)           & $\Delta$                  & Combination ($\%$)                                        \\
\hline
\multirow{2}{*}{$R$(\LbToppipipi)} & \textcolor{white}{0}6.69 $\pm$ 0.33 $\pm$ 0.09 $\pm$ 0.37  & \multirow{2}{*}{$-0.6\sigma$}                    &  \multirow{2}{*}{$\textcolor{white}{0}6.85 \pm 0.19 \pm 0.08 \pm 0.32$}  \\
                                   & \textcolor{white}{0}6.91 $\pm$ 0.23 $\pm$ 0.08 $\pm$ 0.35  &                                                  &                                                          \\
\hline
\multirow{2}{*}{$R($\LbTopKpipi)}  & 16.83 $\pm$  0.49 $\pm$ 0.19 $\pm$ 1.00 & \multirow{2}{*}{$\textcolor{white}{+}1.2\sigma$} &  \multirow{2}{*}{$16.4\textcolor{white}{0} \pm  0.3\textcolor{white}{0} \pm 0.2\textcolor{white}{0} \pm 0.7\textcolor{white}{0}$} \\
                                   & 16.18 $\pm$ 0.33 $\pm$ 0.20 $\pm$ 0.66 &                                                  &                                                          \\
\hline
\multirow{2}{*}{$R($\LbTopKKpi)}   & \textcolor{white}{0}1.14 $\pm$ 0.15 $\pm$ 0.10 $\pm$ 0.16  & \multirow{2}{*}{$-1.4\sigma$}                    &  \multirow{2}{*}{$\textcolor{white}{0}1.32 \pm 0.09 \pm 0.09 \pm 0.10$}  \\
                                   & \textcolor{white}{0}1.39 $\pm$ 0.11 $\pm$ 0.08 $\pm$ 0.10  &                                                  &                                                          \\
\hline
\multirow{2}{*}{$R($\LbTopKKK)}    & \textcolor{white}{0}4.49 $\pm$ 0.22 $\pm$ 0.08 $\pm$ 0.29  & \multirow{2}{*}{$\textcolor{white}{+}2.1\sigma$} &  \multirow{2}{*}{$\textcolor{white}{0}4.11 \pm 0.12 \pm 0.06 \pm 0.19$}  \\
                                   & \textcolor{white}{0}3.97 $\pm$ 0.14 $\pm$ 0.05 $\pm$ 0.20  &                                                  &                                                          \\
\hline
&&&\\
\end{tabular*}
 \end{table}

\begin{table}[t!]
    \setlength{\tabcolsep}{5pt}
   \renewcommand{\arraystretch}{1.2}
\caption{Measurements of the $R$ ratio from the (first row) 2011 and the (second row) 2012 data samples for \Xibz decay modes expressed in per mil as well as their combination. The three uncertainties are statistical, systematic related to the fit model and systematic related to the efficiency, respectively. The consistency of the two determinations for each year, denoted $\Delta$, is quantified as the ratio of the signed difference of the central values over the quadratic sum of the related uncertainties.}

   \label{tab:bf2012}
\newcolumntype{p}[1]{D{p}{\,\pm\,}{#1}}
 \begin{tabular*}{\textwidth}{l p{4.17}  c c  }

      \hline
      $R$ (per decay)       &  \multicolumn{1}{c}{Value $(10^{-3})$}             & $\Delta$                    & Combination $(10^{-3})$ \\
      \hline
      \multirow{2}{*}{$R$(\XibzTopKpipi)} & 7.2\textcolor{white}{0} p 1.4\textcolor{white}{0} \pm 0.2\textcolor{white}{0} \pm  0.9     & \multirow{2}{*}{0.9$\sigma$}  & \multirow{2}{*}{6.2\textcolor{white}{0} $\pm$ 0.8\textcolor{white}{0} $\pm$ 0.2\textcolor{white}{0} $\pm$ 0.8\textcolor{white}{0}}  \\
                                          & 5.8\textcolor{white}{0} p 0.9\textcolor{white}{0} \pm 0.2\textcolor{white}{0} \pm 1.0     &                                &                                                     \\
      \hline
      \multirow{2}{*}{$R$(\XibzTopKpiK)}  & 6.4\textcolor{white}{0} p 1.1\textcolor{white}{0} \pm 0.4\textcolor{white}{0} \pm 0.7    & \multirow{2}{*}{0.9$\sigma$}  &  \multirow{2}{*}{5.6\textcolor{white}{0} $\pm$ 0.6\textcolor{white}{0} $\pm$ 0.4\textcolor{white}{0} $\pm$ 0.5\textcolor{white}{0}}  \\
                                          & 5.3\textcolor{white}{0} p 0.7\textcolor{white}{0} \pm 0.4\textcolor{white}{0} \pm 0.6    &                                &                                                     \\
      \hline
      \multirow{2}{*}{$R$(\XibzTopKKK)} & 0.59 p 0.49 \pm 0.12 \pm 0.10 & \multirow{2}{*}{0.1$\sigma$}  & \multirow{2}{*}{0.57 $\pm$ 0.28 $\pm$ 0.08 $\pm$ 0.10} \\
                                        & 0.56 p 0.34 \pm 0.07 \pm 0.09 &                                &                                                        \\
      \hline
  \end{tabular*}

 \end{table}

\par

As the decay mode \XibzTopKKK is not observed, 90\% and 95\% confidence level (C.L.) intervals, based on the Feldman-Cousins confidence belt inference 
described in Ref.~\cite{Feldman:1997qc}, are placed on the branching fraction for this decay mode relative to  \LbToLcpiLcTopKpi

\begin{center}
$R(\XibzTopKKK) \in [4.05 \mathrm{-} 8.86] \cdot 10^{-4}$ at  90$\%$ C.L., \\
$R(\XibzTopKKK) \in [3.82 \mathrm{-} 9.81] \cdot 10^{-4}$ at  95$\%$ C.L.\textcolor{white}{.} \\
\end{center}

\noindent Using the world-average values $\BF(\LbToLcpi) = (0.430 \pm 0.036) \% $  and $\BF(\LcTopKpi) = (6.46 \pm 0.24) \%$~\cite{HFLAV16}, the branching fractions of the \Lb decay modes are
\begin{alignat*}{6}
&\BF(\LbToppipipi)     &&= (1.90 \pm 0.06 \pm 0.10 \pm 0.16 \pm 0.07) \cdot 10^{-5}&&, \\
&\BF(\LbTopKpipi)       &&= (4.55 \pm 0.08 \pm 0.20 \pm 0.39 \pm 0.17) \cdot 10^{-5}&&, \\
&\BF(\LbTopKKpi)         &&= (0.37 \pm 0.03 \pm 0.04 \pm 0.03 \pm 0.01) \cdot 10^{-5}&&, \\
&\BF(\LbTopKKK)           &&= (1.14 \pm 0.03 \pm 0.07 \pm 0.10 \pm 0.05) \cdot 10^{-5}&&, 
\end{alignat*}              
\noindent where the first uncertainty is statistical and the second comes from experimental systematic sources. The two last uncertainties are due to the knowledge of the branching 
fractions $\BF(\LbToLcpi)$ and $\BF(\LcTopKpi)$ in that order.\\
\indent The product of the branching fractions of the \Xibz decay modes with the hadronisation fraction of \Xibz relative to \Lb are accordingly obtained 
\begin{alignat*}{6}
&\BF(\XibzTopKpipi) \cdot  f_{\Xibz} / f_{\Lb} &&= (1.72 \pm 0.21 \pm 0.25 \pm 0.15 \pm 0.07) \cdot 10^{-6}&&, \\
&\BF(\XibzTopKpiK) \cdot  f_{\Xibz}/ f_{\Lb}  &&= (1.56 \pm 0.16 \pm 0.19 \pm 0.13 \pm 0.06) \cdot 10^{-6}&&, \\
&\BF(\XibzTopKKK) \cdot    f_{\Xibz} / f_{\Lb}     &&\in [0.11 \mathrm{-} 0.25] \cdot 10^{-6}\mbox{ at  90\% C.L.}&&
\end{alignat*}

In summary, the four decay modes \LbTopKpipi, \LbTopKKK, \XibzTopKpipi and \XibzTopKpiK are observed for the first time.
Branching fractions (including the ratio of hadronisation fractions in the case of the \Xibz baryon) of these decay modes and the branching fractions of the two
already observed decay modes \LbToppipipi and \LbTopKKpi~\cite{LHCb-PAPER-2016-030} are determined relative to the \LbToLcpi decay. The \XibzTopKKK decay mode is measured with a significance of $2.3\sigma$ and 90\% and 95\% confidence level intervals are set on its branching fraction relative to \LbToLcpi. The establishment of these signals opens
new channels in which to search for \CP-violating asymmetries in these fully charged four-body decays of \Lb and \Xibz baryons.


%
%
\section*{Acknowledgements}


\noindent We express our gratitude to our colleagues in the CERN
accelerator departments for the excellent performance of the LHC. We
thank the technical and administrative staff at the LHCb
institutes. We acknowledge support from CERN and from the national
agencies: CAPES, CNPq, FAPERJ and FINEP (Brazil); MOST and NSFC
(China); CNRS/IN2P3 (France); BMBF, DFG and MPG (Germany); INFN
(Italy); NWO (The Netherlands); MNiSW and NCN (Poland); MEN/IFA
(Romania); MinES and FASO (Russia); MinECo (Spain); SNSF and SER
(Switzerland); NASU (Ukraine); STFC (United Kingdom); NSF (USA).  We
acknowledge the computing resources that are provided by CERN, IN2P3
(France), KIT and DESY (Germany), INFN (Italy), SURF (The
Netherlands), PIC (Spain), GridPP (United Kingdom), RRCKI and Yandex
LLC (Russia), CSCS (Switzerland), IFIN-HH (Romania), CBPF (Brazil),
PL-GRID (Poland) and OSC (USA). We are indebted to the communities
behind the multiple open-source software packages on which we depend.
Individual groups or members have received support from AvH Foundation
(Germany), EPLANET, Marie Sk\l{}odowska-Curie Actions and ERC
(European Union), ANR, Labex P2IO, ENIGMASS and OCEVU, and R\'{e}gion
Auvergne-Rh\^{o}ne-Alpes (France), RFBR and Yandex LLC (Russia), GVA,
XuntaGal and GENCAT (Spain), Herchel Smith Fund, the Royal Society,
the English-Speaking Union and the Leverhulme Trust (United Kingdom).



\addcontentsline{toc}{section}{References}
\setboolean{inbibliography}{true}
\bibliographystyle{LHCb}
\bibliography{main,LHCb-PAPER,LHCb-CONF,LHCb-DP,LHCb-TDR}

\newpage


 
\newpage
\centerline{\large\bf LHCb collaboration}
\begin{flushleft}
\small
R.~Aaij$^{40}$,
B.~Adeva$^{39}$,
M.~Adinolfi$^{48}$,
Z.~Ajaltouni$^{5}$,
S.~Akar$^{59}$,
J.~Albrecht$^{10}$,
F.~Alessio$^{40}$,
M.~Alexander$^{53}$,
A.~Alfonso~Albero$^{38}$,
S.~Ali$^{43}$
% ...

{\footnotesize \it
% ...
$ ^{5}$Clermont Universit{\'e}, Universit{\'e} Blaise Pascal, CNRS/IN2P3, LPC, Clermont-Ferrand, France\\
% ...
$ ^{10}$Fakult{\"a}t Physik, Technische Universit{\"a}t Dortmund, Dortmund, Germany\\
% ...
$ ^{38}$ICCUB, Universitat de Barcelona, Barcelona, Spain\\
$ ^{39}$Universidad de Santiago de Compostela, Santiago de Compostela, Spain\\
$ ^{40}$European Organization for Nuclear Research (CERN), Geneva, Switzerland\\
% ...
$ ^{43}$Nikhef National Institute for Subatomic Physics, Amsterdam, The Netherlands\\
% ...
$ ^{48}$H.H. Wills Physics Laboratory, University of Bristol, Bristol, United Kingdom\\
% ...
$ ^{53}$School of Physics and Astronomy, University of Glasgow, Glasgow, United Kingdom\\
% ...
$ ^{59}$University of Cincinnati, Cincinnati, OH, United States\\
% ...
}
\end{flushleft}



\end{document}